\documentclass[aps,prd,reprint,superscriptaddress,nofootinbib,longbibliography]{revtex4-2}
\usepackage[T1]{fontenc}
\usepackage{amsmath,amssymb,bm,mathrsfs,mathtools,booktabs}
\usepackage{graphicx,booktabs}
\usepackage[dvipsnames]{xcolor}
\usepackage{hyperref}
\usepackage[normalem]{ulem}
\hypersetup{colorlinks=true,linkcolor=MidnightBlue,citecolor=MidnightBlue,urlcolor=MidnightBlue,pdfauthor={Research manuscript draft},pdftitle={Closed-form rotating black holes beyond Einstein gravity}}
\newcommand{\dd}{\mathrm d}
\newcommand{\Gm}{\Gamma}
\newcommand{\Gb}{\bar\Gamma}

\newcommand{\Pot}{\mathcal P}
\newcommand{\Src}{\mathcal S}
\newcommand{\Real}{\operatorname{Re}}

\newcommand{\Ls}[1]{\mathscr L_{#1}}

\newcommand{\Ts}{\mathsf T}
\newcommand{\Cs}{\mathsf C}
\newcommand{\Js}{\mathsf J}

\newcommand{\Kinv}{\mathcal K}
\newcommand{\Pinv}{\mathcal Q}
\newcommand{\Jinv}{\mathcal J}
\newcommand{\Jodd}{\widetilde{\mathcal J}}
\newcommand{\ellseed}{\hat{\Gamma}}
\newcommand{\Ecal}{\mathcal E}
\newcommand{\Qaux}{Q_{\rm aux}}

\newcommand{\UIUC}{Illinois Center for Advanced Studies of the Universe \& Department of Physics, 
\\ University of Illinois Urbana--Champaign, Urbana, Illinois 61801, USA}

\allowdisplaybreaks
\begin{document}
\title{One Method to Find Them All: \\ Analytic Black Holes beyond Einstein}
\author{Dongjun Li}
\email{dongjun@illinois.edu}
\affiliation{\UIUC}
\author{Jierui Hu}
\email{jieruih2@illinois.edu}
\affiliation{\UIUC}
\author{Nicol\'as Yunes}
\affiliation{\UIUC}
\date{\today}
\begin{abstract}
Rotating black holes offer powerful probes of gravity, but constructing their geometries beyond vacuum general relativity remains challenging. Here we develop a unified framework for constructing stationary, axisymmetric deformations of Kerr from their tidal curvature and physical sources. As a demonstration, we obtain closed-form solutions for six higher-curvature interactions at leading order in the couplings and for generic subextremal spin, without slow-rotation expansions, numerical fitting, or mode decomposition. The solutions pass stringent checks against the field equations, previous mode-by-mode constructions, and known analytic predictions for horizon quantities. We find that the chirality of the interactions organizes the analytic structure of these solutions, a pattern that likely extends to a large class of purely metric higher-curvature theories. Our framework opens a systematic route to black-hole geometries in beyond-Einstein theories and perturbative astrophysical environments. These geometries provide a foundation for strong-field tests with gravitational waves and other observational channels, such as black-hole images and X-ray spectroscopy, as well as theoretical studies of near-extremal black holes.
\end{abstract}
\maketitle

\section*{Introduction}

With nearly 400 binary black hole mergers now cataloged by the LIGO--Virgo--KAGRA Collaboration \cite{LIGOScientific:2026wfs}, black-hole observations have entered an era of increasingly precise tests of strong-field gravity. These opportunities will expand with next-generation ground-based detectors, including Cosmic Explorer \cite{Evans:2021gyd} and Einstein Telescope \cite{Punturo:2010zz}, and space-based missions such as LISA \cite{LISA:2017pwj}, TianQin \cite{TianQin:2015yph}, and Taiji \cite{Ruan:2018tsw}. Complementary electromagnetic probes include black-hole imaging with the Event Horizon Telescope \cite{EventHorizonTelescope:2019dse}, proposed photon-ring measurements with the Black Hole Explorer \cite{Johnson:2024ttr}, and X-ray spectroscopy with Chandra \cite{Weisskopf:2001uu} and NuSTAR \cite{NuSTAR:2013yza}. 

Although current black-hole observations remain consistent with Einstein's general relativity (GR) \cite{LIGOScientific:2021sio, Yunes:2025xwp, Xie:2025voe}, improving precision across these channels offers new opportunities to uncover departures. Such searches are motivated fundamentally by the unresolved reconciliation of GR with quantum mechanics: candidate quantum-gravity frameworks, including string theory \cite{Schwarz:1982jn, Damour:1994zq, Banks:1996vh, Maldacena:1997re, Seiberg:1999vs} and loop quantum gravity \cite{DeWitt:1967yk, Ashtekar:1987gu, Rovelli:1989za, Rovelli:1997yv, Ashtekar:2004eh}, can produce departures from GR in their low-energy descriptions \cite{Donoghue:2012zc}. Cosmological puzzles, including the matter--antimatter asymmetry \cite{Canetti:2012zc}, late-time cosmic acceleration \cite{SupernovaCosmologyProject:1998vns, SupernovaSearchTeam:1998fmf}, and galaxy rotation curves \cite{Sofue:2000jx, Bertone:2016nfn}, further motivate new physics, potentially involving additional fields or modifications of gravity. Connecting these possibilities to black-hole observations requires accurate descriptions of the underlying geometries.

Substantial progress has been made in constructing black holes beyond vacuum GR, spanning high-order slow-rotation expansions \cite{Cano:2019ore}, spectral and pseudospectral treatments of perturbative beyond-GR and environmental corrections \cite{Lam:2025elw, Lam:2025fzi, Fernandes:2025vxg}, with applicability to rapidly rotating and near-extremal configurations \cite{Fernandes:2025vxg, Lam:2026grz}, and some fully nonlinear solutions, including black holes with scalar hair \cite{Herdeiro:2014goa, Garcia:2023ntf, Candan:2026mzh}. Another recent advance is the re-purposing of metric reconstruction as an analytic solver for black-hole geometries beyond GR. Originally developed to recover metric perturbations from curvature perturbations, this technique supports higher-order self-force \cite{Barack:2001ph, Lousto:2002em, Wardell:2021fyy, Barack:2018yvs, Pound:2021qin}, nonlinear ringdown \cite{Campanelli:1998jv, Ma:2024qcv, Khera:2024bjs, Khera:2026lna}, and, more recently, the modeling of environmental \cite{Dyson:2025dlj, Li:2025ffh, Li:2026foy} and beyond-GR \cite{Yang:2026erj, Yang:2026csy} effects on extreme mass-ratio inspirals. In these applications, the Teukolsky equation \cite{Teukolsky:1973ha} determines leading-order curvature perturbations encoding the gravitational-wave strain, while higher-order vacuum-GR or beyond-vacuum-GR sources require the full metric perturbation. 

We recently demonstrated the new role of metric reconstruction by constructing the stationary black hole geometry from curvature sourced by beyond-GR corrections \cite{Hu:2026ute}. This method allowed us to obtain the first mode-by-mode analytic rotating black-hole solution in parity-even cubic gravity at generic subextremal spin. The solution passes stringent field-equation checks and numerically matches independent analytic predictions for horizon quantities \cite{Reall:2019sah}. This advance raises several questions: can the approach accommodate other effective-field-theory (EFT) extensions of GR, including theories with additional fields, as well as perturbative environments? Can the reconstructed metric be expressed in a common form for generic sources? And, more ambitiously, can the mode decomposition itself be avoided?

Here we address these questions by presenting a common quadrature construction for stationary, axisymmetric perturbations of Kerr. Once a conserved effective source is specified, the metric follows from the sourced Teukolsky equation and explicit source quadratures, with homogeneous data fixed by the physical boundary conditions and asymptotic charges. This construction is independent of whether the curvature problem is solved analytically or numerically. We demonstrate its analytic implementation in representative cubic, quartic, and quintic gravity theories, obtaining solutions at generic subextremal spin both mode by mode and, crucially, without a mode decomposition. Using these analytic, closed-form metrics, we calculate representative physical quantities and establish exact analytic agreement with the horizon-thermodynamic predictions of \cite{Reall:2019sah}. We conclude by discussing the broader implications of these results.

\begin{figure*}[p]
    \centering
    \includegraphics[width=\textwidth, height=0.85\textheight, keepaspectratio]{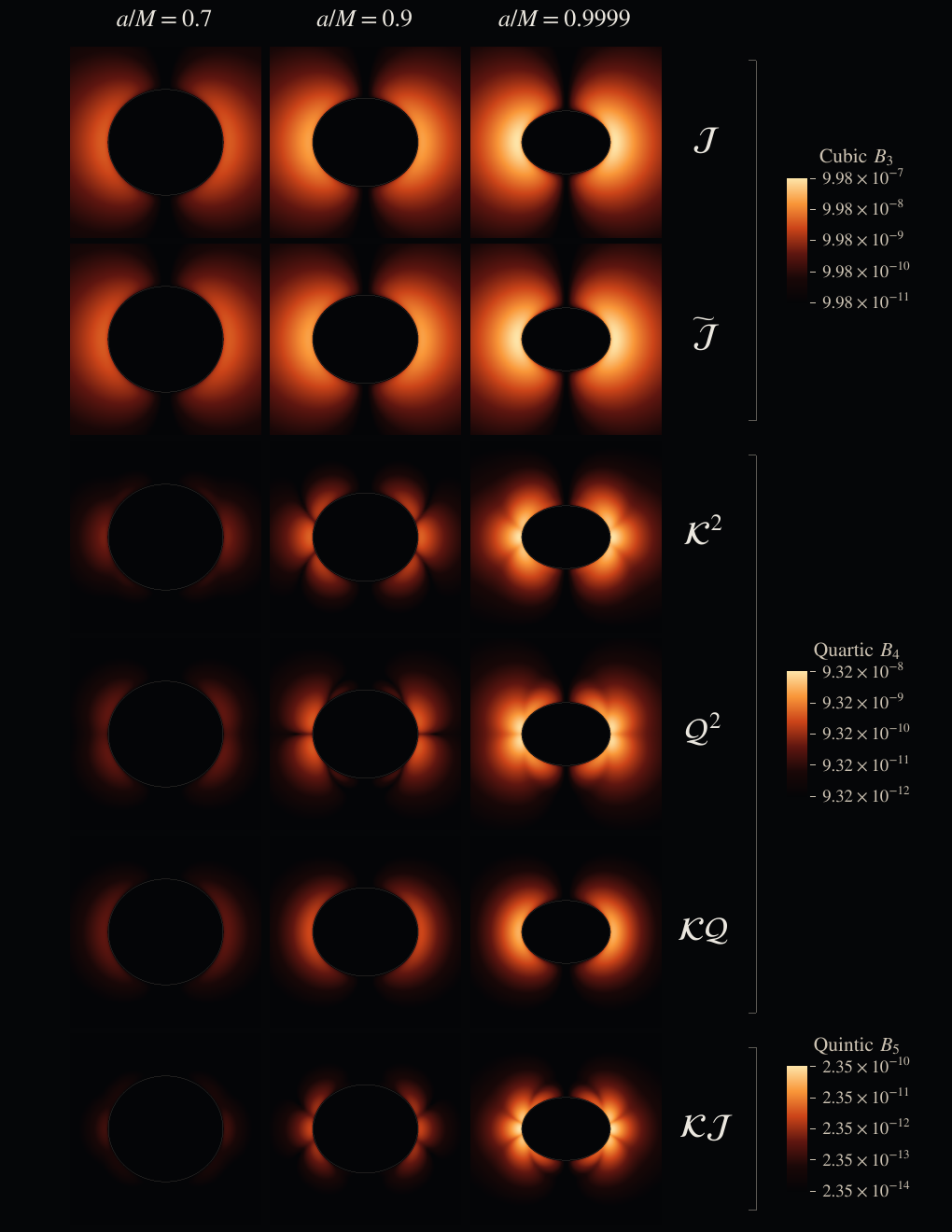}
    \caption{Leading-order Weyl scalar perturbations for six higher-curvature theories. Rows show the cubic interactions $\Jinv$ and $\Jodd$, quartic interactions $\Kinv^2$, $\Pinv^2$, and $\Kinv\Pinv$, and quintic interaction $\Kinv\Jinv$; columns correspond to $a/M=0.7$, $0.9$, and $0.9999$. Colors show the dimensionless transverse-Weyl amplitude $\mathcal B_n$. Each interaction is considered separately, with dimensionless coupling $\lambda_I=(\ell_c/M)^{2n_I-2}$, where $n_I$ is its curvature order and $\ell_c/M=0.03$. Four-decade logarithmic color scales are shared across all panels of the same curvature order. All panels use the same mass and spatial scale. Black regions mask $r<r_++0.03M$, where $r_+$ is the outer-horizon radius of the Kerr background. The two cubic amplitudes coincide at each spin because their Weyl perturbations $\Psi_0^{(1)}$ differ only by an overall phase, although their metric corrections $h_{\mu\nu}^{(1)}$ are distinct. Across the sampled spins, $\mathcal B_n$ generally increases toward extremality.}
    \label{fig:Weyl_scalars}
\end{figure*}

\section*{General reconstruction of rotating black holes beyond vacuum GR}
\label{sec:generic}

Consider a stationary, axisymmetric deformation $g_{\mu\nu}=g^{(0)}_{\mu\nu}+\epsilon h_{\mu\nu}^{(1)}+O(\epsilon^2)$ of a Kerr black hole with mass $M$ and spin parameter $a$, where $|a|<M$ and $\epsilon$ is a small dimensionless bookkeeping parameter that can be defined in terms of beyond-GR coupling constants. We use the superscript $(1)$ to denote any linear-in-$\epsilon$ perturbations of the Kerr background. At the retained order, we assume that the metric perturbation $h_{\mu\nu}^{(1)}$ satisfies the linearized field equations:
\begin{equation} \label{eq:linear_EE}
    \delta{G}_{\mu\nu}[h^{(1)}]=\Src_{\mu\nu}[g^{(0)}]\,,\quad
    \nabla^\mu\Src_{\mu\nu}=0\,,
\end{equation}
where $\delta{G}_{\mu\nu}$ is the linearized Einstein tensor, and $\Src_{\mu\nu}=8\pi\tilde{T}_{\mu\nu}$, with $\tilde{T}_{\mu\nu}$ the normalized stress-energy tensor driven by either the beyond-GR correction or matter in the environment, with the perturbation parameter $\epsilon$ factored out. In other words, the stress-energy tensor always enters at $\mathcal{O}(\epsilon)$, as the zeroth-order background is the Kerr metric. When $\tilde{T}_{\mu\nu}$ is coupled to extra fields (e.g.~in certain scalar-tensor theories \cite{Alexander:2009tp, Sotiriou:2013qea}), the equations for these fields must first be solved consistently at their required perturbative order. Henceforth, we focus on the case for which $\Src_{\mu\nu}$ only involves the metric, and extend our treatment to cases with extra non-metric fields in the Discussion. 

Unlike most other approaches which solve the linearized field equations in Eq.~\eqref{eq:linear_EE} directly, we follow the strategy in~\cite{Li:2026rkf, Hu:2026ute} to solve instead the sourced, stationary Teukolsky equations directly, i.e.,
\begin{align} \label{eq:teuk}
    \mathcal{H}_{i}\Psi_{i}^{(1)}=\mathcal{S}_{i}\,,\quad
    i\in\{0,4\}\,,
\end{align}
where $\Psi_{0,4}$ are the time-independent Weyl scalars in the Newman--Penrose formalism \cite{Newman:1961qr} characterizing non-radiative, transverse tidal curvature along the principal null directions, as explained below. For the Kerr background, $\Psi_{0,4}^{(0)}=0$ while beyond-GR corrections and matter can induce nonzero corrections $\Psi_{0,4}^{(1)}$. The operators $\mathcal{H}_{0,4}$ are the standard Teukolsky operators in \cite{Teukolsky:1973ha}, while $\mathcal{S}_{0,4}$ are the Teukolsky sources generated by $\mathcal{S}_{\mu\nu}$, as defined in \cite{Teukolsky:1973ha, Li:2026rkf}. We showed in Ref.~\cite{Hu:2026ute} that stationary, axisymmetric perturbations preserving the simultaneous reflection $(t,\phi)\mapsto(-t,-\phi)$ satisfy $\Psi_4^{(1)}=\Delta^2\Psi_0^{(1)}/(4\Gb^4)$, where $\Delta=r^2-2Mr+a^2$ and $\Gm=r+ia\cos\theta$ in Boyer--Lindquist coordinates $(t,r,\theta,\phi)$, with the overbar denoting complex conjugation. Thus, it is sufficient to solve one of $\Psi_{0,4}^{(1)}$ and obtain the other via the above identity.

We obtain the metric perturbation $h_{ab}^{(1)}$ from the Weyl scalars  $\Psi_{0,4}^{(1)}$ via the sourced, metric reconstruction scheme that we developed in~\cite{Li:2026rkf}, where a hierarchy of Newman--Penrose equations is solved systematically. Generalizing the parity-even cubic construction of~\cite{Hu:2026ute}, we obtain the following quadrature representation for conserved sources and boundary data preserving the reflection symmetry $(t,\phi)\mapsto(-t,-\phi)$, including all six higher-curvature solutions considered below:
\begin{align}
    & h_{ab}^{(1)}
    =\mathscr{R}_{ab}[\Pot]+\mathscr{Q}_{ab}[\Src]
    +h^{\rm comp}_{ab}\,, \label{eq:schematic}\\
    & \Pot(r,\theta)
    =\frac{1}{6}\int_r^\infty
    (u-r)^3\Psi_0^{(1)}(u,\theta)\,du\,,\nonumber
\end{align}
where $a$ and $b$ denote indices of the background tetrad $e_{a}^{\mu}$, which we here choose to be the Kinnersley tetrad \cite{Teukolsky:1973ha}. 
For sources without reflection symmetry, the black hole metric is still given by Eq.~\eqref{eq:schematic}, but with $4\Gb^4\Psi_4^{(1)}/\Delta^2$ in place of $\Psi_0^{(1)}$ in the master potential $\Pot$. The fixed, second-order differential operator $\mathscr{R}_{ab}$ acting on $\Pot$ can be extracted from Eq.~(11) in \cite{Hu:2026ute} and is universal across different theories and sources. The integrals of the $\mathcal{S}_{\mu\nu}$ source, $\mathscr{Q}_{ab}$, also have a universal form, though the results of these integrals vary from theory to theory. The last term $h^{\rm comp}_{ab}$ contains homogeneous charge and gauge data, selected by the physical boundary conditions. Note that when the spin-two $\Psi_{0}^{(1)}$ vanishes, Eq.~\eqref{eq:schematic} still retains source-driven, monopole and dipole information through $\mathscr{Q}_{ab}$, as discussed in detail in \cite{Li:2026rkf}. The explicit forms of $\mathscr{R}_{ab}$ and $\mathscr{Q}_{ab}$ are given in the Methods section, with a general prescription for how to fix $h^{\rm comp}_{ab}$. 

The general structure in Eq.~\eqref{eq:schematic} is particularly convenient for constructing black hole geometries that perturbatively deviate from the Kerr spacetime, as one does not need any longer to repeatedly solve the transport equations theory by theory as in~\cite{Li:2026rkf, Hu:2026ute}. The complexity of computing $h_{ab}^{(1)}$ has been largely reduced to: (i) solving the stationary Teukolsky equations in Eq.~\eqref{eq:teuk}, which can be done using a Green's function; (ii) evaluating the quadratures given by $\mathscr{Q}_{ab}[\Src]$; and (iii) determining $h^{\rm comp}_{ab}$ from proper boundary conditions. The difficulty is much smaller than directly solving the coupled tensorial Einstein equations of~\eqref{eq:linear_EE}. Furthermore, when solving for $\Psi_{0}^{(1)}$ and evaluating $\mathscr{Q}_{ab}[\Src]$, one can freely choose different strategies, including analytical approaches (with or without mode decomposition), numerical approaches (spectral, pseudospectral, or finite difference, etc.), or a combination. In obtaining Eq.~\eqref{eq:schematic}, other sourced, metric reconstruction schemes might also be viable \cite{Aksteiner:2016pjt, Green:2019nam, Wardell:2024yoi}, but some of them \cite{Aksteiner:2016pjt, Wardell:2024yoi} are not directly valid in the stationary limit, so those methods would first need to be extended for this purpose, though they should reach the same answer as Eq.~\eqref{eq:schematic} under the same gauge and homogeneous data.

\section*{Weyl scalars and tidal deformation}
\label{sec:tidal}

Before applying Eq.~\eqref{eq:schematic} to specific theories, we clarify the physical information carried by $\Psi_{0,4}^{(1)}$. For time-dependent perturbations, the gauge-invariant Weyl scalars $\Psi_{0,4}^{(1)}$ characterize gravitational radiation crossing the horizon and reaching null infinity \cite{Teukolsky:1973ha}. For stationary perturbations, they remain gauge invariant~\cite{Hu:2026ute} but instead encode non-radiative transverse tidal curvature. Consider a narrow bundle of light rays with affinely parametrized tangent $k^\mu$, and two orthonormal transverse vectors $e_A^\mu$, $A=1,2$, spanning its local cross section. The optical tidal matrix
\begin{equation} \label{eq:optical}
    \mathcal T_{AB}[k]
    =R_{\mu\alpha\nu\beta}e_A^\mu k^\alpha e_B^\nu k^\beta
    =\mathcal{W}_{AB}[k]+\frac12R_{kk}\delta_{AB}
\end{equation}
controls the relative transverse acceleration of neighboring rays, where $\mathcal W_{AB}[k]\equiv
C_{\mu\alpha\nu\beta}e_A^\mu k^\alpha e_B^\nu k^\beta$ is the trace-free Weyl contribution, stretching the bundle in one transverse direction while compressing it in the other, and $R_{kk}\equiv R_{\mu\nu}k^\mu k^\nu$ determines the isotropic Ricci contribution.

Consider a null tetrad $e^{\mu}_{a}=\{l^{\mu},n^{\mu},m^{\mu},\bar{m}^{\mu}\}$ in the Newman--Penrose formalism, where $l^{\mu}$ and $n^{\mu}$ are the principal outgoing and ingoing null directions, respectively, and the angular span $\{m^{\mu},\bar{m}^{\mu}\}$ is fixed by orthogonality conditions \cite{Newman:1961qr, Chandrasekhar:1985kt}. Choosing $e_1^{\mu}=(m^\mu+\bar{m}^{\mu})/\sqrt{2}$ and $e_2^{\mu}=-i(m^\mu-\bar{m}^{\mu})/\sqrt{2}$, we obtain the 2-dimensional matrices
\begin{align} \label{eq:tidal-matrices}
    \mathcal{T}_{AB}[l]
    &=
    \begin{pmatrix}
        \operatorname{Re}\Psi_0 & \operatorname{Im}\Psi_0\\
        \operatorname{Im}\Psi_0 &-\operatorname{Re}\Psi_0
    \end{pmatrix}
    + \begin{pmatrix}
        \frac12R_{ll} & 0\\
        0 &\frac12R_{ll}
    \end{pmatrix}
    \,,\nonumber\\
    \mathcal{T}_{AB}[n]
    &=
    \begin{pmatrix}
        \operatorname{Re}\Psi_4 &-\operatorname{Im}\Psi_4\\
        -\operatorname{Im}\Psi_4&-\operatorname{Re}\Psi_4
    \end{pmatrix}
    + \begin{pmatrix}
        \frac12R_{nn} &0\\
        0&\frac12R_{nn}
    \end{pmatrix}\,.
\end{align}
For the Kerr spacetime in vacuum, the two principal null directions $l^{\mu}$ and $n^{\mu}$ are special, as they are shear free, i.e., $\mathcal{T}_{AB}[l]=0=\mathcal{T}_{AB}[n]$ or, equivalently, $\Psi_0^{(0)}=0=\Psi_4^{(0)}$. A circular cross section along the congruences of these null rays can expand and rotate but maintains its circular shape without turning into an ellipse at any time. This fact, however, does not imply that the Kerr spacetime is free of tidal forces. For example, in Kerr, zero-angular-momentum infalling null rays generally follow nonprincipal directions $k^\mu$ and experience nonzero transverse tides, $\mathcal T_{AB}[k]\ne0$~\cite{Lam:2026grz}. Equatorial principal null rays instead satisfy $L_z=aE$, where $E$ is their conserved energy. For the deformed black hole geometries considered here, the nonzero stationary $\Psi_{0,4}^{(1)}$ then measure the anisotropic tidal curvature introduced along null directions continuously connected to Kerr's principal null directions.

\section*{Analytic black holes in higher-curvature gravity}
\label{sec:higher_derivative}

For purely metric higher-curvature gravity, the reconstruction in Eq.~\eqref{eq:schematic} leads to an analytic spacetime metric without mode decomposition. Our earlier parity-even cubic solution expressed $\mathscr Q_{ab}[\Src]$ through rational functions of $\Gm=r+ia\cos{\theta}$ and $\Gb$ in Boyer--Lindquist coordinates, while determining the curvature potential $\Pot$ via mode decomposition \cite{Hu:2026ute}. Here, we extend the construction to a broad set of purely metric higher-curvature gravity theories and eliminate the mode decomposition of $\Pot$, with concrete results provided for six cubic, quartic, and quintic interactions. The key to this breakthrough is a common generating kernel that exploits the rational structure of the curvature sources, yielding closed-form solutions at generic subextremal spin.

We consider the four-dimensional action
\begin{align}\label{eq:action}
    S={}&\frac{1}{16\pi G}\int d^4x\sqrt{-g}
    \big[R+\alpha_{\rm e}\Jinv+\alpha_{\rm o}\Jodd
    +\beta_1\Kinv^2+\beta_2\Pinv^2 \nonumber\\
    & \qquad+\beta_3\Kinv\Pinv+\gamma_1\Kinv\Jinv
    +\mathcal{F}(\Kinv,\Pinv,\Jinv,\Jodd)\big]\,,
\end{align}
where $\Kinv=R_{\alpha\beta\mu\nu}R^{\alpha\beta\mu\nu}$, $\Pinv=R_{\alpha\beta\mu\nu}\tilde{R}^{\alpha\beta\mu\nu}$, $\Jinv=R_{\alpha\beta}{}^{\mu\nu}R_{\mu\nu}{}^{\rho\sigma}R_{\rho\sigma}{}^{\alpha\beta}$, and $\Jodd=R_{\alpha\beta}{}^{\mu\nu}R_{\mu\nu}{}^{\rho\sigma}\tilde R_{\rho\sigma}{}^{\alpha\beta}$, with $R_{\alpha\beta\mu\nu}$ the Riemann tensor and its dual $\tilde{R}_{\alpha\beta\mu\nu}
=\epsilon_{\mu\nu\rho\sigma}R^{\rho\sigma}{}_{\alpha\beta}/2$. The coupling constants $c_I = \{\alpha_{\rm e,o},\beta_i,\gamma_1\}$ have dimensions of length to the fourth, sixth, and eighth powers, respectively. Thus, to parameterize interactions of different curvature orders uniformly, we introduce a common characteristic length scale $\ell_c$ and set  $c_I=\ell_c^{2n_I-2}$ for each interaction $I$ at curvature order $n_I$, such that the coupling combination $\lambda_I=c_I/M^{2n_I-2} = (\ell_c/M)^{2n_I-2}$ is dimensionless and we identify $\lambda_I$ with the perturbation parameter $\epsilon$. The two cubic and three quartic interactions cover the leading vacuum higher-curvature corrections to Einstein gravity through eight derivatives in the EFT expansion, modulo field redefinitions and topological terms \cite{Endlich:2017tqa, Cano:2019ore, Reall:2019sah}, while $\Kinv\Jinv$ represents one of four independent algebraic quintic interactions \cite{Ruhdorfer:2019qmk}. The function $\mathcal{F}$ collects the remaining quintic and higher-order algebraic-curvature terms. Covariant derivatives of curvature and couplings to non-metric fields can extend this action further.

Chirality provides a useful organizing principle for these interactions. In four dimensions, the Weyl tensor separates into self-dual and anti-self-dual components, $C^\pm=(C\mp i\tilde C)/2$, where $\tilde{C}_{\alpha\beta\mu\nu}=\epsilon_{\mu\nu\rho\sigma}C^{\rho\sigma}{}_{\alpha\beta}/2$ is its Hodge dual. These components have definite chirality, defined by the Hodge-duality
condition $\widetilde{C}^{\pm}=\pm iC^{\pm}$, and are exchanged by parity, $(\hat P C)^\pm = \hat P(C^\mp)$, where $\hat P$ denotes spatial inversion
acting on tensor fields~\cite{Ruhdorfer:2019qmk}. Single-chirality terms involve only one of $C^{\pm}$, whereas mixed-chirality terms involve both. A real interaction can combine separate single-chirality terms, so this classification does not by itself determine parity. On a Ricci-flat background, $\Jinv$, $\Jodd$, $\Kinv^2-\Pinv^2$, and $\Kinv\Pinv$ contain only separate single-chirality contributions, whereas $\Kinv^2+\Pinv^2$ couples the sectors, and $\Kinv\Jinv$ contains both types.

This classification further organizes the analytic structure of both the sources and the solutions of the Teukolsky equations [i.e., Eq.~\eqref{eq:teuk}] and the Einstein equations [i.e., Eq.~\eqref{eq:linear_EE}]. In the Kinnersley tetrad, the only nonzero background Weyl scalar is $\Psi_2=-M/\Gb^3$, and Ricci flatness implies that the background Riemann and Weyl tensors coincide. Varying the interactions in Eq.~\eqref{eq:action} and evaluating the resulting sources in the field equations on Kerr therefore produces effective stress-energy tensors $\mathcal{S}_{\mu\nu}$ [i.e., Eq.~\eqref{eq:linear_EE}] whose components are rational functions in $\Gm$ and $\Gb$, after known coordinate and tetrad factors are separated. The differential operators generating the Teukolsky source $\mathcal{S}_0$ in Eq.~\eqref{eq:teuk} preserve this structure, up to explicit spin--boost factors.

For the Teukolsky source $\mathcal{S}_0$ studied here, single-chirality contributions only have poles in either $\Gm$ or $\Gb$, whereas mixed-chirality contributions have poles in both. After transforming the sourced spin-$2$ Teukolsky equation in Eq.~\eqref{eq:teuk} to an auxiliary spin-$0$ scalar problem, we find a common generating kernel to solve these families of sourced equations directly, without resolving individual angular modes. The single-chirality contributions to $\Psi_0^{(1)}$ require only rational functions, square roots, and logarithms, while the mixed-chirality contributions admit finite combinations of multivariate hypergeometric functions, with equivalent elliptic-iterated representations. 

The source-driven metric correction $\mathscr{Q}_{ab}[\Src]$ has a different analytic structure. After explicit tetrad factors are removed, this correction is composed of rational functions only for all the single-chirality contributions, $\Jinv$, $\Jodd$, $\Kinv^2-\Pinv^2$, and $\Kinv\Pinv$. The mixed-chirality quartic combination $\Kinv^2+\Pinv^2$ and the quintic interaction $\Kinv\Jinv$ additionally introduce $\log(\Gm/\Gb)$ terms with rational function coefficients. Consequently, the individual parity-even quartic interactions, $\Kinv^2$ and $\Pinv^2$, have both rational and logarithmic dependence. The analytic structure of the deformed metric and the chiral classification, in terms of rational functions and logarithms or hypergeometric functions, extend beyond the theories we considered, as long as the interactions are built from the Riemann tensor and its dual. In the Methods, we detail the auxiliary scalar problem and generating-kernel construction, while the Supplemental Information provides the explicit solutions of $\Psi_0^{(1)}$ and $h_{ab}^{(1)}$ for all six higher-curvature interactions in Eq.~\eqref{eq:action}. 

For all six interactions, we verify that the metric deformations in Eq.~\eqref{eq:schematic} satisfy the linearized field equations in Eq.~\eqref{eq:linear_EE} and that the Weyl perturbations $\Psi_0^{(1)}$ computed from these metrics satisfy the sourced Teukolsky equations in Eq.~\eqref{eq:teuk} exactly. The solutions are regular at the horizon and decay at infinity for subextremal spins, and also agree with mode-by-mode constructions following Ref.~\cite{Hu:2026ute}. For example, at $a/M=0.9999$, comparisons over five polar angles and $200$ radii logarithmically spaced in $r-r_+$, from $r=r_++10^{-6}M$ to $r=100M$, yield a maximum relative discrepancy in $\Psi_0^{(1)}$ below $1.4\times10^{-22}$ across all six theories with the mode sums truncated at $\ell_{\max}=100$. The corresponding \texttt{Mathematica} notebooks are provided in Ref.~\cite{bh_solutions}.

The corrected black-hole geometries retain Kerr's stationarity and axisymmetry, with Killing vectors $(\partial_t)^\mu$ and $(\partial_\phi)^\mu$, and, by construction, the same ADM mass $M$ and angular momentum $J=aM$ as the background. For each theory with fixed coupling $c_I$, the perturbative family selected by our boundary conditions is therefore labeled by $M$ and $J$. The radial coordinate is asymptotically areal: the area $\mathcal{A}(r)$ of a constant-$t$, constant-$r$ surface satisfies $\mathcal{A}(r)/(4\pi r^2)\to1$ as $r\to\infty$. Despite these similarities, their surface gravity, horizon angular velocity $\omega_+$, shadow contours, and extremality bounds can differ from Kerr's, as examined in the next section. In our chosen coordinates and gauge, the horizon generally shifts away from the background location $r_+=M+\sqrt{M^2-a^2}$, and this displacement is included consistently when evaluating its properties. Distinct from this coordinate-dependent displacement, the horizon area $\mathcal{A}_{\mathcal H}$ can also differ from their Kerr values at fixed $M$ and $J$. For the parity-odd interactions, however, the first-order area correction vanishes by reflection symmetry, even though the horizon geometry can be deformed \cite{Cano:2019ore}. Although the stationary and axial Killing vectors retain their coordinate form, the horizon-generating combination $\xi^\mu=(\partial_t)^\mu+\omega_+(\partial_\phi)^\mu$ can change through the correction to $\omega_+$.

\begin{figure*}[p]
    \centering
    \includegraphics[width=\textwidth, height=0.825\textheight, keepaspectratio]{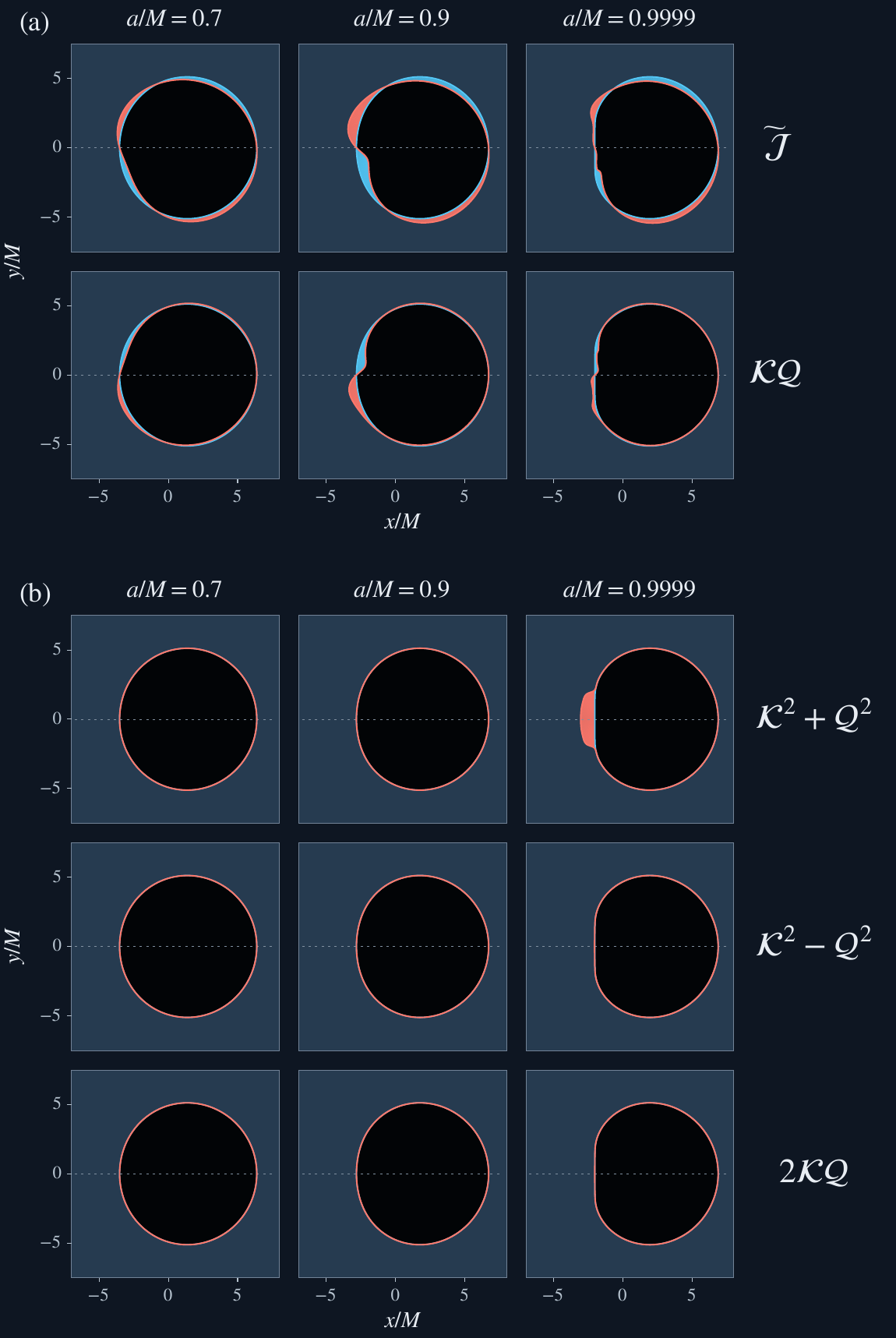}
    \caption{Parity and chirality in black-hole shadow deformations. (a) Parity-odd interactions $\Jodd$ and $\Kinv\Pinv$, with displacement magnifications $A=3\times10^6$ and $5\times10^8$, respectively. The first-order displacement is antisymmetric under reflection across the horizontal screen axis, allowing the contour to deform without changing its area at first order. (b) Quartic responses in the chiral basis $\Kinv^2+\Pinv^2$, $\Kinv^2-\Pinv^2$, and $2\Kinv\Pinv$, all with $A=2\times10^5$. The first combination is mixed-chirality, which dominates over the latter two single-chirality contributions. In both panels, columns correspond to $a/M=0.7$, $0.9$, and $0.9999$, respectively. Each indicated interaction or combination is considered separately, with dimensionless coupling $\lambda_I=(\ell_c/M)^{2n_I-2}$ and $\ell_c/M=0.03$, where $n_I$ is its curvature order. Blue curves denote Kerr, and red curves show $\rho_{\rm Kerr}+A\,\delta\rho$. Red and blue shading indicate outward and inward radial displacement, with black filling the Kerr interior. Both couplings and magnification factors are fixed across spins; only the first-order displacement is magnified. The observer is static at $r_{\rm obs}=50M$ and inclination $i=90^\circ$, with the same screen convention throughout.}
    \label{fig:shadow_parity_chirality}
\end{figure*}

\section*{Physical properties and observables of rotating modified black holes}
\label{sec:observables}

We begin by examining transverse tidal distortions around the corrected black holes. As shown in Eq.~\eqref{eq:tidal-matrices}, the real and imaginary parts of $\Psi_0^{(1)}$ describe the local stretching and squeezing of narrow light bundles along the outgoing null direction connected to Kerr's principal direction, while $\Psi_4^{(1)}$ describes the corresponding ingoing response. To compare their combined strength without fixing the relative normalization of these null directions, Fig.~\ref{fig:Weyl_scalars} displays the dimensionless transverse-Weyl amplitude $\mathcal{B}_n=|\lambda_I|M^2\sqrt{|\Psi_0^{(1)}\Psi_4^{(1)}|}
=|\lambda_I|M^2\Delta|\Psi_0^{(1)}|/(2\Gm\Gb)$ for interaction $I$. This combination is invariant under spin--boost transformations, and the second expression follows from the reflection symmetry between $\Psi_0^{(1)}$ and $\Psi_4^{(1)}$. We show $\mathcal B_n$ across spins $\chi=J/M^2=a/M\in\{0.7,0.9,0.9999\}$ at fixed $\ell_c/M=0.03$.\footnote{Higher-curvature interactions generally shift the extremality bound away from the Kerr value $\chi_{\rm ext}=1$ \cite{Reall:2019sah}. At the couplings used here, $|\chi_{\rm ext}-1|<1.2\times10^{-7}$ (Supplemental Information), much smaller than $1-\chi=10^{-4}$ for our highest-spin configuration. The $\chi=0.9999$ panels therefore remain near-extremal with respect to the corrected bound.} Across the sampled spins, $\mathcal B_n$ generally increases toward Kerr extremality in all six theories, indicating an enhancement of this transverse-curvature response.

At fixed ADM mass and angular momentum, the surface gravity and horizon angular velocity provide invariant benchmarks. Our earlier mode-by-mode solution in parity-even cubic gravity \cite{Hu:2026ute} reproduced independent thermodynamic predictions in \cite{Reall:2019sah} numerically. With the analytic, mode-free solutions here, we now establish exact analytic agreement for the first-order corrections to these horizon quantities in all cubic and quartic interactions for which the thermodynamic results are available \cite{Reall:2019sah}. The derivations are given in the Methods and Supplemental Information, with accompanying \texttt{Mathematica} notebooks \cite{bh_solutions}.

As a demonstration of how observables can be computed from our analytic metrics, we compute the first-order photon capture boundaries, also known as critical curves, which serve as a proxy for black hole shadows. These boundaries distinguish captured from escaping rays on an observer's screen \cite{Gralla:2019xty}. Assuming minimally coupled photons in geometric optics, we integrate the null geodesic equations using ray tracing, without assuming geodesic separability, since the modified black holes do not necessarily possess a Kerr-like Carter constant, as they can be Petrov type I \cite{Owen:2021eez}. The observer is static at $r_{\rm obs}=50M$ in the equatorial plane, and we consider $a/M\in\{0.7,0.9,0.9999\}$. We compute the corresponding first-order corrections to the shadow contours using the procedure described in the Methods.

\begin{figure*}[p]
    \centering
    \includegraphics[width=\textwidth, height=0.9\textheight, keepaspectratio]{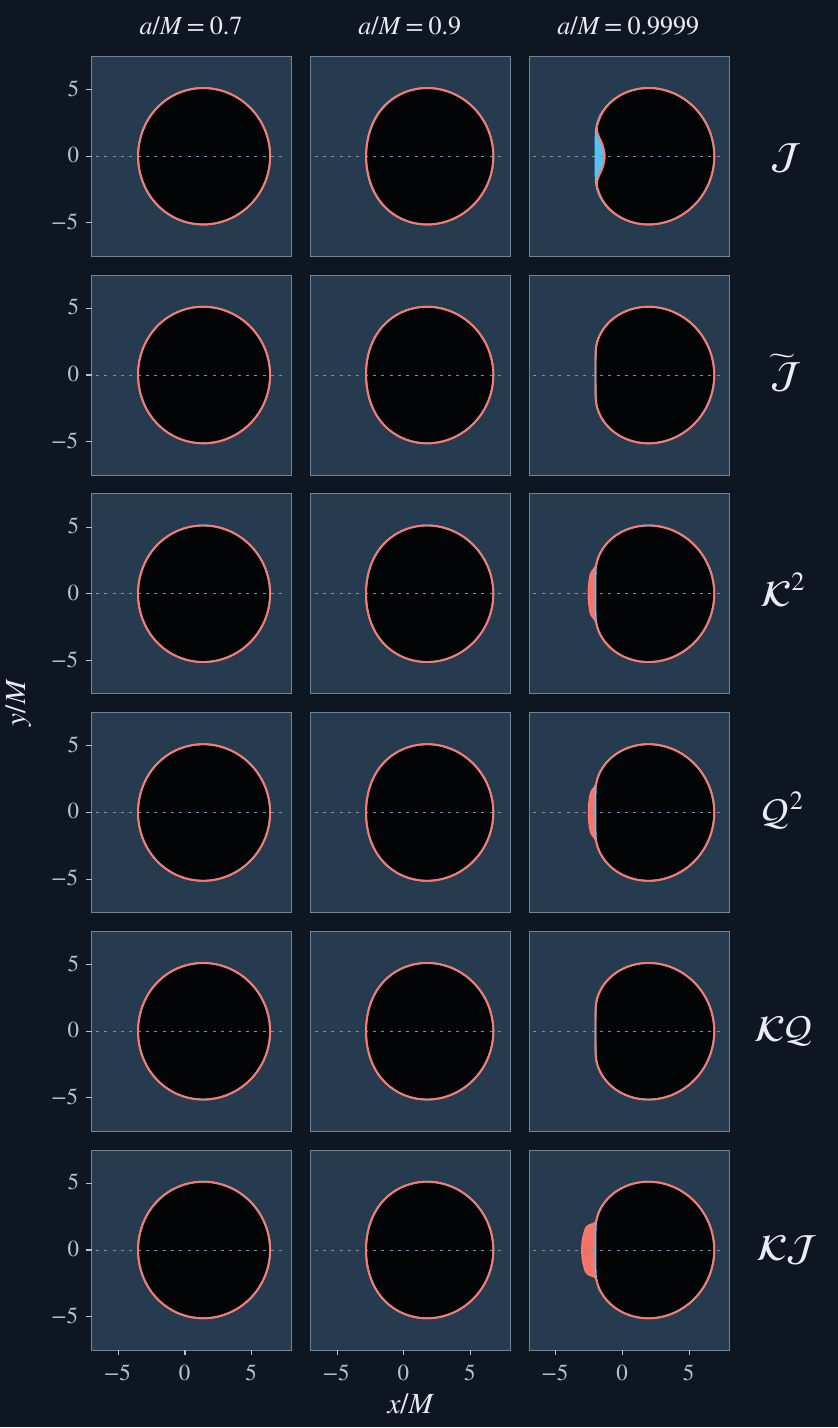}
    \caption{Equatorial shadow responses for the six higher-curvature interactions, considered separately at $\ell_c/M=0.03$, with the same spins, observer configuration, and color conventions as Fig.~\ref{fig:shadow_parity_chirality}. The first-order displacement is magnified by $A=5\times10^4$, $2\times10^5$, and $5\times10^8$ for cubic, quartic, and quintic interactions, respectively, with each factor fixed across spins.}
    \label{fig:shadow_comparison}
\end{figure*}

Figures~\ref{fig:shadow_parity_chirality} and \ref{fig:shadow_comparison} illustrate the resulting silhouettes and compare their contours with Kerr critical curves. The magnitude of the corrections is typically small for $\ell_c/M = 0.03$, so, to make the small corrections visible in the figure, we magnify only the first-order radial displacement, $\rho_{\rm shown}=\rho_{\rm Kerr}+A\,\delta\rho$, with magnification factors $A$ fixed across spins within each curvature order.\footnote{Within the first-order approximation, the same displayed contours could be obtained with unit magnification by replacing $\lambda_I$ with $\lambda_I^{\rm eff}=A\lambda_I$, corresponding to $\ell_{c,I}^{\rm eff}=A^{1/(2n_I-2)}\ell_c$. In Fig.~\ref{fig:shadow_comparison}, the effective couplings $\lambda_I^{\rm eff}$ are $4.05\times10^{-2}$, $1.458\times10^{-4}$, and $3.2805\times10^{-4}$ for the cubic, quartic, and quintic interactions, respectively. The quartic combinations in Fig.~\ref{fig:shadow_parity_chirality}(b) likewise have $\lambda_I^{\rm eff}=1.458\times10^{-4}$. For the more strongly magnified parity-odd responses in Fig.~\ref{fig:shadow_parity_chirality}(a), the effective couplings $\lambda_I^{\rm eff}$ are $2.43$ and $0.3645$ for $\Jodd$ and $\Kinv\Pinv$, respectively. Although $\lambda_{\Jodd}^{\rm eff}=2.43>1$, this value merely expresses the visual amplification as an effective coupling; the physical coupling remains $\lambda_{\Jodd}=8.1\times10^{-7}$. The display magnification therefore does not alter the underlying small-coupling expansion or itself indicate a breakdown of perturbation theory.} For most interactions, the deviations increase toward extremality across the sampled spins at fixed coupling. The parity-odd interactions $\Jodd$ and $\Kinv\Pinv$, however, exhibit a non-monotonic spin dependence: the more strongly magnified contours in the upper panel of Fig.~\ref{fig:shadow_parity_chirality} show their largest displacement at $a/M=0.9$, with a smaller response at $a/M=0.9999$. Thus, approaching extremality does not universally amplify beyond-GR signatures.

The parity of higher-curvature interactions leaves a particularly clear imprint on the shadow. For equatorial viewing in the upper panel of Fig.~\ref{fig:shadow_parity_chirality}, the first-order contour displacement induced by the parity-odd interactions $\Jodd$ and $\Kinv\Pinv$ is antisymmetric under reflection across the horizontal screen axis, whereas the parity-even response is symmetric. Odd-parity corrections can therefore shift or distort the contour and break its top--bottom symmetry without changing the shadow area at first order, because the area changes at reflected points cancel. This optical asymmetry is consistent with the equatorial-reflection breaking previously demonstrated in rotating black-hole geometries with parity-violating higher-curvature interactions \cite{Cardoso:2018ptl, Cano:2019ore}.

The quartic shadow responses reveal additional structure in the chiral basis. The lower panel of Fig.~\ref{fig:shadow_parity_chirality} compares $\Kinv^2+\Pinv^2$, $\Kinv^2-\Pinv^2$, and $2\Kinv\Pinv$. The first combination is mixed-chirality, whereas the latter two contain separate single-chirality contributions. The displayed contours suggest substantially larger deformations in the mixed-chiral combination than in either single-chiral combination, indicating that the individual $\Kinv^2$ and $\Pinv^2$ responses are dominated by their shared mixed-chirality sector. This pattern suggests that the chiral structure organizing the analytic solutions also shapes their optical signatures. Combined with emission and photon-propagation modeling, these geometries provide a foundation for testing fundamental symmetries of beyond-GR black holes with the Event Horizon Telescope \cite{EventHorizonTelescope:2019dse} and proposed Black Hole Explorer \cite{Johnson:2024ttr}.

\section*{Discussion}

We have established a unified construction of rotating black-hole geometries beyond vacuum GR at leading perturbative order. Once a perturbative problem is expressed as a conserved source on Kerr, the metric follows from a curvature solution, universal source quadratures, and physical completion data. Applying this approach to purely metric higher-curvature gravity theories built from the Riemann tensor and its dual, we discover an analytic route to construct rotating black-hole solutions without slow-rotation expansions, post-Minkowskian expansions, numerical fitting, or mode decomposition. For the six cubic, quartic, and quintic interactions computed as a demonstration, our solutions retain the full radial, angular, and spin dependence at generic subextremal spin and pass stringent checks, including reproducing the independent analytic predictions for horizon quantities obtained from black hole thermodynamics \cite{Reall:2019sah}.

Beyond supplying metrics in specific theories, our framework reveals why distinct higher-curvature interactions admit a common analytic treatment. The relation between the stationary spin-two equation and an auxiliary scalar geometry organizes rational sources around a shared generating kernel, with chirality distinguishing the analytic function classes that arise. Related chiral structures enter parity isospectrality and its breaking in ringdown \cite{Weller:2026jqu}, and may provide a geometric explanation for the decoupling of Kerr's Weyl-scalar equations \cite{Green:2026nlt}. Our shadow calculations reveal a further observation: shadow deformations for equatorial viewing are predominantly driven by the mixed-chirality sector for quartic interactions. The analytic metrics also provide direct access to near-extremal geometry, where higher-curvature effects can be enhanced \cite{Horowitz:2023xyl, Horowitz:2024dch, Lam:2026grz}. Although the validity of our rotating black hole spacetimes at generic subextremal spin does not guarantee regularity or perturbative control in the extremal limit, these analytic metrics allow us to investigate the approach to extremality analytically and identify physical quantities that diagnose the breakdown of the EFT description.

Our framework also extends beyond purely metric theories. Once additional fields or matter configurations are determined at the required perturbative order, their conserved effective sources enter the same metric construction framework, Eq.~\eqref{eq:schematic}. Natural applications include theories coupled to additional non-metric fields, such as dynamical Chern--Simons and scalar Gauss--Bonnet gravity \cite{Alexander:2009tp, Sotiriou:2013qea}, as well as astrophysical environments ranging from ultralight scalar clouds, for which we have obtained results using this approach \cite{Lam:2026def}, to dark-matter distributions and accretion disks. Depending on source complexity, further applications may admit mode-free solutions, require mode-by-mode analytic constructions \cite{Hu:2026ute}, or rely on spectral and other numerical methods. Each strategy feeds into the same metric construction framework, whose applicability therefore does not depend on closed-form solvability. Time-dependent environments offer a further extension, replacing stationary radial quadratures with characteristic transport and dynamical boundary data.

Observationally, these geometries connect the physical origin of black-hole deformations to their measurable signatures. Combined with well-developed techniques, such as the modified Teukolsky formalism \cite{Li:2022pcy, Hussain:2022ins, Cano:2023tmv, Li:2025fci, Yang:2026csy, Li:2026foy}, formulated in the same Newman--Penrose framework, and other spectral approaches \cite{Chung:2024ira, Lui:2026osx}, they provide the basis for ringdown and extreme mass-ratio inspiral predictions. Orbital and photon-propagation models further connect the metrics to black-hole imaging, as demonstrated here, and X-ray spectroscopy. Treating beyond-GR and environmental backreaction within a shared framework provides a consistent basis for assessing their competing signatures and combining constraints across these channels.

The broader opportunity these rotating black hole metrics provide is to replace isolated, theory-specific constructions with a systematic map from physical sources to black-hole geometries and observables. Extending this map across beyond-GR theories and astrophysical environments can reveal how analytic structure, spacetime symmetries, and near-extremal behavior shape observable strong-field phenomena. Together with increasingly precise gravitational-wave and electromagnetic observations, this framework can turn black holes into precision laboratories for testing the limits of general relativity, characterizing the surrounding matter or dark matter, and exploring the fundamental structure of spacetime.

\section*{Acknowledgements}
We are grateful to Nicholas Loutrel for fruitful discussions during the early stages of this project. We thank Roman Berens and Kelvin Lam for insightful discussions. D.L., J.H., and N.Y. acknowledge support from the Simons Foundation (via Award No. 896696), the Simons Foundation International (via Grant No. SFI-MPS-BH-00012593-01), and the NSF (via Grant No. PHY-2512423). Codex (OpenAI) and Claude Code (Anthropic) were used to assist with code development and debugging, consistency checks, figure preparation, and English grammar corrections. All AI-assisted outputs have been independently verified by the authors.
\bibliographystyle{apsrev4-2}
\bibliography{references}
\clearpage
\onecolumngrid
\section*{Methods}
\subsection*{Conventions, notations, and charge-fixing}

Throughout this work, we use signature $(-,+,+,+)$ and the Newman--Penrose conventions of~\cite{Chandrasekhar:1985kt,Hu:2026ute}. The background Kinnersley tetrad in Boyer--Lindquist coordinates is
\begin{equation} \label{eq:method-tetrad}
    l^\mu=\frac{(r^2+a^2,\Delta,0,a)}{\Delta}\,,\qquad
    n^\mu=\frac{(r^2+a^2,-\Delta,0,a)}{2\Sigma}\,,\qquad
    m^\mu=\frac{(ia\sin\theta,0,1,i/\sin\theta)}{\sqrt2\Gm}\,,
\end{equation}
with $l_{\mu} n^{\mu}=-1$ and $m_{\mu}\bar m^{\mu}=1$. We set
\begin{align} \label{eq:method-notation}
    x&=\cos\theta\,,\qquad
    w=1-x^2\,,\qquad
    \Gm=r+iax\,,\qquad 
    \Gb=r-iax\,,\qquad \nonumber\\
    \Sigma
    &=\Gm\Gb\,,\qquad 
    \Delta=r^2-2Mr+a^2\,,\qquad 
    j=ia\sin\theta\,,
\end{align}
and define the following angular operator and background Newman–Penrose quantities in Kerr
\begin{equation} \label{eq:reconstruction-abbreviations}
    \Ls{k}=\partial_\theta+k\cot\theta\,,\quad
    \mu^{(0)}=-\frac{\Delta}{2\Gm\Gb^2}\,,\quad
    \pi^{(0)}=\frac{j}{\sqrt2\Gb^2}\,,\quad
    \tau^{(0)}=-\frac{j}{\sqrt2\Sigma}\,,\quad
    \Psi_2^{(0)}=-\frac{M}{\Gb^3}\,.
\end{equation}

For each higher-curvature interaction $I$, we expand
\begin{equation}\label{eq:coupling-normalization}
    g_{\mu\nu}
    =g^{(0)}_{\mu\nu}
    +\epsilon h^{I(1)}_{\mu\nu}
    +\mathcal{O}(\epsilon^2)\,,\qquad
    \epsilon=\lambda_I\equiv\frac{c_I}{M^{2n_I-2}}\,,
\end{equation}
where $c_I=(\alpha_{\rm e},\alpha_{\rm o},\beta_1,\beta_2,
\beta_3,\gamma_1)$ are the dimensionful couplings in Eq.~\eqref{eq:action},
and $n_I=(3,3,4,4,4,5)$. Thus,
\begin{equation}\label{eq:dimensionless-couplings}
    \lambda_I=
    \left(\frac{\alpha_{\rm e}}{M^4},
    \frac{\alpha_{\rm o}}{M^4},\frac{\beta_1}{M^6},
    \frac{\beta_2}{M^6},\frac{\beta_3}{M^6},
    \frac{\gamma_1}{M^8}\right)\,.
\end{equation}
All first-order fields, sources, potentials, and metric
corrections are deformations per unit $\lambda_I$, with the
interaction label suppressed when unambiguous.

The equation of motion at first order in $\epsilon$ is $\delta{G}_{\mu\nu}[h^{(1)}]=\Src_{\mu\nu}[g^{(0)}]$, where $\Src_{\mu\nu}=8\pi\tilde{T}_{\mu\nu}$ is the stationary, axisymmetric source driven by either the beyond-GR correction or matter in the environment. The corresponding Newman--Penrose Ricci scalars used in the reconstruction hierarchy of \cite{Li:2026rkf, Hu:2026ute} are
\begin{equation} \label{eq:method-Ricci}
    \Phi_{22}=\tfrac12\Src_{nn}\,,\quad
    \Phi_{21}=\tfrac12\Src_{n\bar m}\,,\quad
    \Phi_{20}=\tfrac12\Src_{\bar m\bar m}\,,\quad
    \Phi_{11}=\tfrac14(\Src_{ln}+\Src_{m\bar m})\,,\quad
    \Phi_{12}=\bar{\Phi}_{21}\,,\quad
    \Lambda=-\tfrac1{24}g^{\mu\nu(0)}\Src_{\mu\nu}\,,
\end{equation}
where indices with tetrad vector symbols represent contractions with the tetrad, e.g., $\Src_{nn}=\Src_{\mu\nu}n^{\mu}n^{\nu}$.

For the higher-curvature solutions considered here, we impose $\mathcal S_{\hat a\hat b}=\mathcal O(r^{-5})$ or faster and $h_{\hat a\hat b}^{(1)}=\mathcal O(r^{-3})$ near spatial infinity and in an asymptotically inertial orthonormal frame, labeled by hatted lowercase Latin indices. The metric falloff fixes the ADM mass and angular-momentum shifts, while the remaining homogeneous data are selected by horizon and axis regularity and the residual-gauge prescription. The condition $h=\mathcal O(r^{-3})$ as one approaches spatial infinity sets the shifts of the ADM mass and angular momentum to zero, since such shifts contribute to the metric in the orthonormal frame at $\mathcal{O}(r^{-1})$ and $\mathcal{O}(r^{-2})$, respectively. A different choice of charge requires
\begin{equation} \label{eq:completion}
    h^{\rm comp}_{\mu\nu}=\delta M\,\partial_M g^{(0)}_{\mu\nu}\big|_a
    +\delta a\,\partial_a g^{(0)}_{\mu\nu}\big|_M
    +\mathcal L_\xi g^{(0)}_{\mu\nu}\,,\qquad 
    \delta a=\frac{\delta J_{\rm ADM}-a\delta M}{M}\,,
\end{equation}
with $\xi$ restoring the chosen gauge.

\subsection*{The reconstructed metric}

We now give the curvature and source contributions $\mathscr R_{ab}[\Pot]$ and $\mathscr Q_{ab}[\Src]$ in Eq.~\eqref{eq:schematic}. In the traceful outgoing radiation gauge, $h_{\mu\nu}^{(1)}n^\nu=0$, the metric components are solved from the transport hierarchy in \cite{Li:2026rkf, Hu:2026ute}, where
\begin{subequations}\label{eq:metric}
\begin{align}
    & h_{m\bar m}^{(1)}=\Ts\,, \\
    & h_{\bar m\bar m}^{(1)}
    =-\frac{2\Gm}{\Gb^2}(\Gm\partial_r^2-2\partial_r)\Pot\,,\\
    & h_{l\bar m}^{(1)}
    =\frac{2\sqrt2\Gm}{\Gb^2\Delta}
    [\Sigma\Ls2\partial_r-2r\Ls2-j(\Gm-\Gb)\partial_r]\Pot+\Cs\,,\\
    & h_{ll}^{(1)}
    =\frac{\Sigma^2}{\Delta^2}
    \left[-8\Real\left\{\frac1{\Gb^2}\left(\Ls1+\frac{2j}{\Gm}\right)\Ls2\Pot\right\}+\Js\right]\,.
\end{align}
\end{subequations}
The terms containing $\Pot$ define $\mathscr R_{ab}$ in the metric, which is the contribution from Weyl scalars $\Psi_{0,4}^{(1)}$. The two real functions $\Ts,\Js$ and the complex function $\Cs$ define $\mathscr Q_{ab}$. These functions are radial quadratures of the Newman--Penrose Ricci scalars $\Phi_{ij}$, $i,j\in\{0,1,2\}$, and $\Lambda$. For higher-curvature gravity, the closed-form analytic expressions for $\Ts$, $\Js$, and $\Cs$ can be found in the Supplemental Information.

The Weyl scalar perturbation $\Psi_0^{(1)}$ satisfies the stationary Teukolsky equation
\begin{equation} \label{eq:H0}
    \mathcal{H}_0\Psi_0^{(1)}
    =\mathcal{S}_0\,,\qquad
    \mathcal{H}_0
    =\Delta\partial_r^2+6(r-M)\partial_r
    +w\partial_x^2-2x\partial_x+2-\frac{4x^2}{w}\,.
\end{equation}
The source $\mathcal{S}_0$ is obtained from the Ricci tensor as in Refs.~\cite{Teukolsky:1973ha,Li:2026rkf}. The perturbed Weyl scalar $\Psi_0^{(1)}$ can be solved analytically or numerically using the Green's function~\cite{Linet:1977vx,Hu:2026ute} associated with the operator $\mathcal{H}_0$, which can be uniquely determined by imposing horizon regularity and decay at infinity.  For the six higher-curvature gravity theories, $\Psi_0^{(1)}$ has closed-form, analytic expressions, which are shown in the next section.

We use the reflection-symmetric specialization described in the main text, for which $4\Gb^4\Psi_4^{(1)}/\Delta^2=\Psi_0^{(1)}$. Let us define the radial moments at fixed $\theta$ by
\begin{equation} \label{eq:reconstruction-tail}
    \mathcal{I}_n[f](r,\theta)
    =\frac1{n!}\int_r^\infty(u-r)^n f(u,\theta)du\,,\qquad 
    \Pot=\mathcal{I}_3[\Psi_0^{(1)}]\,,
\end{equation}
where the function $f(r,\theta)$ inside $\mathcal{I}_n$ is evaluated at $u$ instead of $r$. Then, the source-driven parts $\Ts$, $\Cs$, and $\Js$ of $h_{ab}^{(1)}$ in Eq.~\eqref{eq:metric} are
\begin{align} \label{sm:quadratures}
    & \Ts
    =-8\Real\left\{\frac1{\Gb}\mathcal I_1
    \left[\frac{\Sigma^2\Gb}{\Delta^2}\Phi_{22}\right]\right\}\,,\qquad
    \Cs
    =\frac{2\Gm}{\Delta}\mathcal I_0
    \left[\Gb(2v_s-\bar\tau^{(0)}\Ts)\right]\,,\nonumber\\
    & \Js
    =\frac{2\Delta}{\Sigma}\Ts
    -8\Real\left\{\Gb m_s+\frac2{\Gb}\mathcal{I}_1
    \left[\frac{\Sigma\Gb^2}{\Delta}\mathcal F_s\right]\right\}
    -\mathcal{I}_0[\mathcal B_0]\,,
\end{align}
where $m_s$, $U$, and $v_s$ are the source-driven contributions to $\mu^{(1)}$, $\Psi_3^{(1)}$, and $\pi^{(1)}$, respectively:
\begin{align}
    & m_s
    =-\frac{\Delta}{4\Sigma}
    \left(\partial_r+\frac1\Gm-\frac1\Gb\right)\Ts\,,\qquad
    U
    =\frac{2\Delta}{\Gb^6}
    \mathcal I_0\left[\frac{\Sigma\Gb^6}{\Delta^2}R_h\right]\,, \nonumber\\
    & v_s
    =-\frac{2\Gm}{\Gb}\mathcal I_0
    \left[\frac{\Gb^2}{\Delta}\left\{U+\Phi_{21}
    +\left(\pi^{(0)}+\bar{\tau}^{(0)}\right)m_s
    +\mu^{(0)}\bar{\tau}^{(0)}\Ts\right\}\right]\,.
\end{align}
The additional auxiliary functions $\mathcal{F}_s$ and $\mathcal{B}_0$ are
\begin{align} \label{sm:B}
    \mathcal{F}_s
    =& \;-3\Psi_2^{(0)}m_s+
    \left(\frac1{\sqrt2\Gm}\Ls1-2\tau^{(0)}\right)U+R_g\,,\nonumber\\
    \mathcal{B}_0
    =& \;8\Real\left[\left(1-\frac\Gb\Gm\right)m_s\right]
    +4\sqrt2\Real\left[\frac\Gb\Gm\Ls1v_s-\frac j\Gm v_s\right]
    -\frac{2\sqrt2\Delta}{\Sigma}\Real
    \left[j\left(\frac2{\Gm^2}+\frac1\Sigma\right)\Cs\right] \nonumber\\
    & \;+\frac{2ra^2w-4M(r^2-a^2x^2)}{\Sigma^2}\Ts+16r\Lambda\,,
\end{align}
where $R_h$ and $R_g$ are the sources entering the transport equations of $\Psi_3^{(1)}$ [i.e., $\mathcal{R}_h^{(1)}$ in Eq.~(17)] and $\Psi_2^{(1)}$ [i.e., $\mathcal{R}_g^{(1)}$ in Eq.~(21)] in \cite{Li:2026rkf}, respectively:
\begin{align} \label{eq:reconstruction-Rg}
    & R_h
    =\left(- \frac{\Delta}{2\Sigma}\partial_r+2\mu^{(0)}+2\bar{\mu}^{(0)}
    +\frac{r-M}{\Sigma}\right)\Phi_{21}
    -\left(\frac1{\sqrt2\Gb}\partial_\theta+2\pi^{(0)}
    -\bar{\tau}^{(0)}\right)\Phi_{22}\,,\nonumber\\
    & R_g
    =\left(\frac1{\sqrt2\Gm}\Ls1+2\bar{\pi}^{(0)}\right)\Phi_{21}
    -\left(\partial_r+\frac1\Gm\right)\Phi_{22}
    +\frac{\Delta}{\Sigma}\partial_r\Lambda
    -2\mu^{(0)}\Phi_{11}+2\pi^{(0)}\Phi_{12}\,.
\end{align}

The full metric perturbation in the coordinate basis is
\begin{equation} \label{eq:assemble}
    h_{\mu\nu}^{(1)}
    =h_{ll}^{(1)}n_\mu n_\nu
    -4\Real\left[h_{l\bar{m}}^{(1)}n_{(\mu}m_{\nu)}\right]
    +2\Real\left[h_{\bar{m}\bar{m}}^{(1)}m_\mu m_\nu\right]
    +2h_{m\bar{m}}^{(1)} m_{(\mu}\bar m_{\nu)}\,,
\end{equation}
where parentheses denote symmetrization with weight $1/2$. The factors of $\Delta^{-1}$ and $\Delta^{-2}$ in the Kinnersley tetrad do not by themselves diagnose a singular metric, as these tetrad vectors are multiplied by non-trivial functions of $\Delta$ as well. For each source considered in this work, we have explicitly checked horizon regularity in an ingoing chart using $h_{m\bar{m}}^{(1)}$, $h_{\bar{m}\bar{m}}^{(1)}$, $\Delta h_{l\bar{m}}^{(1)}/\Sigma$, and $\Delta^2h_{ll}^{(1)}/\Sigma^2$, and we have found no angular divergence.

\subsection*{Generating kernel and the master potentials}
\label{meth:generator}

As discussed earlier, the interactions in Eq.~\eqref{eq:action} can be classified by chirality. After decomposing the Weyl tensor $C$ into two chiral sectors $C^\pm=(C\mp i\tilde{C})/2$, the single-chirality contributions contract curvature tensors within either sector separately, whereas mixed-chirality contributions couple both sectors. On the Ricci-flat Kerr background, defining 
\begin{equation}
    \mathcal{I}_\pm
    =\Kinv\pm i\Pinv
    =2C^{\mp}_{\alpha\beta\mu\nu}C^{\mp \alpha\beta\mu\nu}\,,\qquad
    \mathcal J_\pm
    =\Jinv\pm i\Jodd
    =2C^{\mp}_{\alpha\beta}{}^{\mu\nu}C^{\mp}_{\mu\nu}{}^{\rho\sigma}
    C^{\mp}_{\rho\sigma}{}^{\alpha\beta}\,,
\end{equation}
we find
\begin{equation} \label{eq:chiral-quartics}
    \Jinv=\frac{\mathcal J_++\mathcal J_-}{2}\,,\qquad
    \Jodd=\frac{\mathcal J_+-\mathcal J_-}{2i}\,,\qquad
    \Kinv^2-\Pinv^2
    =\frac{\mathcal{I}_+^2+\mathcal{I}_-^2}{2}\,,\qquad
    \Kinv\Pinv
    =\frac{\mathcal{I}_+^2-\mathcal{I}_-^2}{4i}\,,
\end{equation}
which contain only separate single-chirality contributions. On the other hand,
\begin{equation}
    \Kinv^2+\Pinv^2
    =\mathcal{I}_{+}\mathcal{I}_{-}\,,\qquad
    \Kinv\Jinv
    =\frac{\mathcal{I}_{+}\mathcal{J}_{+}
    +\mathcal{I}_{-}\mathcal{J}_{-}}{4}
    +\frac{\mathcal{I}_{+}\mathcal{J}_{-}
    +\mathcal{I}_{-}\mathcal{J}_{+}}{4}\,,
\end{equation}
where $\Kinv^2+\Pinv^2$ is mixed-chirality, and $\Kinv\Jinv$ contains both the single-chirality and mixed-chirality contributions. If we write down the remaining quintic interactions, we will find that $\Kinv\Jinv-\Pinv\Jodd$ contains only single-chirality contributions, whereas $\Kinv\Jinv+\Pinv\Jodd$ is mixed-chirality.

This chirality classification has helped us divide up the Teukolsky source $\mathcal{S}_{0}$ into two sectors, where the single-chirality sector $\mathcal{S}^{\rm ch}_{n}$ at each curvature order $n$ has a pole either in $\Gb$ or $\Gm$, i.e.,
\begin{align} \label{eq:single_chirality_sources}
    \mathcal{S}^{\rm ch}_{3}
    =-43200M^6a^2w\,\Gb\Gm^{-11}\,,\quad
    \mathcal{S}^{\rm ch}_{4}
    =-3953664M^9a^2w\,\Gb\Gm^{-14}\,,\quad
    \mathcal{S}^{\rm ch}_{5}
    =-31933440M^{12}a^2w\Gb\Gm^{-17}\,,
\end{align}
while the mixed-chirality sector $\mathcal{S}^{\rm mix}_{n}$ at each curvature order $n$ has mixed poles in $\Gb$ and $\Gm$, i.e.,
\begin{align} \label{eq:mixed_chirality_sources}
    & \mathcal{S}^{\rm mix}_{4}
    =-967680M^{9}a^2w\Gm^{-4}\Gb^{-9}\,,\nonumber\\
    & \mathcal{S}^{\rm mix}_{5}
    =-207360M^{12}a^2w\left(22\Gm^{-4}\Gb^{-12}+7\Gm^{-7}\Gb^{-9}
    -6\Gm^{-9}\Gb^{-7}+5\Gm^{-11}\Gb^{-5}\right)\,,
\end{align}
such that the Teukolsky sources $\mathcal{S}_{I}$ for each higher-curvature interaction $I$ in Eq.~\eqref{eq:action} are
\begin{equation} \label{eq:Teukolsky_sources_all}
    \mathcal{S}_{\Jinv}=\mathcal{S}^{\rm ch}_{3}\,,\quad
    \mathcal{S}_{\Jodd}=-i\mathcal{S}^{\rm ch}_{3}\,,\quad
    \mathcal{S}_{\Kinv^2}
    =\mathcal{S}^{\rm mix}_{4}+\mathcal{S}^{\rm ch}_{4}\,,\quad
    \mathcal{S}_{\Pinv^2}
    =\mathcal{S}^{\rm mix}_{4}-\mathcal{S}^{\rm ch}_{4}\,,\quad
    \mathcal{S}_{\Kinv\Pinv}=-i\mathcal{S}^{\rm ch}_{4}\,,\quad
    \mathcal{S}_{\Kinv\Jinv}
    =\mathcal{S}^{\rm mix}_{5}+\mathcal{S}^{\rm ch}_{5}\,.
\end{equation}
We will see that this source structure also manifests in the solutions for $\Psi_0^{(1)}$ and the master potential $\Pot$.

To solve the sourced Teukolsky equation in Eq.~\eqref{eq:teuk}, we find it helpful to transform the equation into an auxiliary scalar equation. Defining $b=\sqrt{M^2-a^2}$ and $y=r-M$, we find
\begin{equation} \label{eq:intertwiner}
    \mathcal{H}_0\mathscr{D}=\mathscr{D}L_0\,,\qquad
    \mathscr{D}=w\partial_y^2\partial_x^2\,,\qquad
    L_0=\partial_y(\Delta\partial_y)+\partial_x(w\partial_x)\,,
\end{equation}
where $L_0=\Gm\Gb\Box_{\rm Kerr}$ on stationary, axisymmetric scalars, with $\Box_{\rm Kerr}$ the scalar-wave operator on Kerr. Thus, if $\mathcal{S}_0=\mathscr{D}S$ and $L_0V=S$, then $\Psi_0^{(1)}=\mathscr{D}V$. We then transform the problem of solving Eq.~\eqref{eq:teuk} into determining the auxiliary source $S$ and solving the auxiliary scalar equation $L_0V=S$. Furthermore, if we perform an additional prolate-coordinate map $\varrho=\sqrt{\Delta w}$ and $Z=yx$, the auxiliary scalar operator $L_0$ becomes a flat-space Poisson operator
\begin{equation} \label{eq:auxiliary-geometry}
    L_0=(y^2-b^2x^2)\nabla_{\mathbb R^3}^{2}\,,
\end{equation}
where $\nabla_{\mathbb{R}^3}^{2}$ is the Laplacian on the auxiliary Euclidean space $\mathbb{R}^3$. Under such a map, the Kerr horizon maps to the line segment $\varrho=0$, $|Z|\le b$.

To solve the auxiliary scalar equation $L_0V=S$, with $S$ determined by the Teukolsky sources in the form of Eqs.~\eqref{eq:single_chirality_sources}--\eqref{eq:Teukolsky_sources_all}, we introduce a generating function $F$ satisfying
\begin{equation} \label{eq:seed}
    F(y,x;\mathfrak{m},\mathfrak{a})=-\int_0^1 H_s^{-1}\,ds\,,\qquad
    L_0F=\ellseed^{-2}\,,
\end{equation}
where 
\begin{equation} \label{eq:kernel-polynomials}
    H_s=B_s(y+b+sd)+\mathfrak{a}sT_s(x-1)\,,\quad
    B_s=\mathfrak{m}+b+\mathfrak{a}(2s-1)\,,\quad
    T_s=2b+sd\,,\quad
    d=\mathfrak{m}+\mathfrak{a}-b\,,\quad
\end{equation}
and $0\le s\le1$. In this case, $\Gm$ and $\Gb$ can be represented as $\ellseed=y+\mathfrak{m}+\mathfrak{a}x$, evaluated at $\mathfrak{m}=M$ and $\mathfrak{a}=\pm ia$, respectively. The auxiliary parameters $\mathfrak{m}$ and $\mathfrak{a}$ vary the source, while the Kerr background stays fixed. Because $L_0$ is independent of $\mathfrak{m}$ and $\mathfrak{a}$, differentiating $F$ generates higher poles in Eqs.~\eqref{eq:single_chirality_sources}--\eqref{eq:Teukolsky_sources_all} and their solutions together:
\begin{equation} \label{eq:source-generation}
    L_0\left(\partial_\mathfrak{m}^m\partial_\mathfrak{a}^nF\right)
    =(-1)^{m+n}(m+n+1)!\frac{x^n}{\ellseed^{m+n+2}}\,,
\end{equation}
where the derivatives in $\mathfrak{m}$ and $\mathfrak{a}$ are taken at fixed $M,a,y,x$ before substitution.

The potential in the reconstruction is $\Pot=\mathcal I_3[\Psi_0^{(1)}]$, which requires an additional radial integral over $F$.
It is therefore convenient to define the radial integral
\begin{align}
 \Ecal(y,x;\mathfrak{m},\mathfrak{a})
 &=\mathcal I_1[\partial_x^2F]
 =\int_y^\infty(v-y)\,\partial_x^2F(v,x;\mathfrak{m},\mathfrak{a})\dd v,\nonumber\\
 \partial_y^2\Ecal&=\partial_x^2F,\qquad
 \partial_y^4(w\Ecal)=\mathscr D F.
 \label{eq:potential-generator}
\end{align}
The derivatives $\partial_\mathfrak{a}$ and $\partial_\mathfrak{m}$ that generate $\Psi_0^{(1)}$ from $F$ now generate $\Pot$ from $w\Ecal$. The analytic expressions for $F$ and $\Ecal$ are found to be
\begin{align}
    & F(y,x;\mathfrak{m},\mathfrak{a})
    =\frac1{2\mathfrak{a}\sqrt{\Qaux}}
    \left[2\log\frac{A_F+\sqrt{\Qaux}}{E_F+\sqrt{\Qaux}}
    +\log\frac{(b+\zeta)(y+b)}{\ellseed(\mathfrak{m}+\mathfrak{a}+\zeta)}\right]\,, \\
    & \Ecal(y,x;\mathfrak{m},\mathfrak{a})
    =\frac1{w^2}\left[(Z_*^2+\Qaux)F
    +\frac{A_*}{2\mathfrak{a}}\log
    \frac{\mathfrak{m}+b+\mathfrak{a}}{\mathfrak{m}+b-\mathfrak{a}}
    -\frac{Z_*}{\mathfrak{a}}\log\frac{\ellseed}{y+b}\right]
    +\frac{(\mathfrak{m}-b)x-\mathfrak{a}}{2\mathfrak{a}w}\,,
\end{align}
where
\begin{align}
    & A_F
    =\ellseed+\mathfrak{m}x+\mathfrak{a}+\zeta\,,\qquad 
    E_F=y+bx+b+\zeta\,,\qquad
    Z_*=xy-\zeta\,,\qquad 
    A_*=y(1+x^2)-\zeta x(3-x^2)\,, \nonumber\\
    & \Qaux
    =y^2+b^2x^2-2\zeta yx+\zeta^2-b^2\,,\qquad 
    \zeta
    =\frac{\mathfrak{m}^2-\mathfrak{a}^2-b^2}{2\mathfrak{a}}\,.
\end{align}

Writing $[f]_*=[f]_{\mathfrak{m}=M,\mathfrak{a}=ia}$ and $[f]_t=[f]_{\mathfrak{m}=M,\mathfrak{a}=ia(2t-1)}$, we find that the single-chirality contributions to $\Pot$ due to the sources in Eq.~\eqref{eq:single_chirality_sources} are
\begin{align} \label{eq:single_chirality_potential}
    \Pot^{\rm ch}_{3}
    =\frac{M^6w}{42}
    \left[(3+\mathfrak{a}\partial_\mathfrak{a})
    \partial_\mathfrak{m}^4\Ecal\right]_*\,,\quad
    \Pot^{\rm ch}_{4}
    =-\frac{M^9w}{1575}
    \left[(9+2\mathfrak{a}\partial_\mathfrak{a})
    \partial_\mathfrak{m}^7\Ecal\right]_*\,,\quad
    \Pot^{\rm ch}_{5}
    =\frac{M^{12}w}{655200}
    \big[(12+2\mathfrak{a}\partial_\mathfrak{a})
    \partial_\mathfrak{m}^{10}\Ecal\big]_*\,,
\end{align}
such that
\begin{equation}
    \Pot_{\Jinv}=\Pot^{\rm ch}_{3}\,,\qquad
    \Pot_{\Jodd}=-i\Pot^{\rm ch}_{3}\,,\qquad
    \Pot_{\Kinv\Pinv}=-i\Pot^{\rm ch}_{4}\,,
\end{equation}
which follows the same structure as Eq.~\eqref{eq:Teukolsky_sources_all}. The potential $\Pot^{\rm ch}_{n}$ is a shared single-chirality contribution at curvature order $n$.

For the mixed poles $\Gm^{-m}\Gb^{-n}$, with $m,n\ge1$, in the mixed-chirality sources of Eq.~\eqref{eq:mixed_chirality_sources}, we use the following Feynman-parameter identity to rewrite $\Gm^{-m}\Gb^{-n}$:
\begin{equation} \label{eq:beta-poles}
    \Gm^{-m}\Gb^{-n}
    =\frac{(N-1)!}{(m-1)!(n-1)!}\int_0^1
    \frac{t^{m-1}(1-t)^{n-1}}{[r+ia(2t-1)x]^N}\,dt\,,
\end{equation}
where $N=m+n$ and $N\ge6$ for all the interactions in Eq.~\eqref{eq:action}. Since the integrand in Eq.~\eqref{eq:beta-poles} is proportional to $\hat{\Gm}^{-N}|_t=\hat{\Gm}^{-N}|_{\mathfrak{m}=M,\mathfrak{a}=ia(2t-1)}$, the potential $\Pot$ from the mixed-chirality source can also be generated from $\Ecal$, with an additional integral over $t$ by linearity. Thus, the mixed-chirality potentials are found to be:
\begin{align} \label{eq:mixed_chirality_potential}
    & \Pot^{\rm mix}_{4}
    =4M^9a^2w\int_0^1t^3(1-t)^8
    \left[\partial_\mathfrak{m}^7
    \left(\frac{\Ecal}{\mathfrak{a}^2}\right)\right]_t\,dt\,,\qquad
    \Pot^{\rm mix}_{5}
    =M^{12}a^2w\int_0^1W_5(t)
    \left[\partial_\mathfrak{m}^{10}
    \left(\frac{\Ecal}{\mathfrak{a}^2}\right)\right]_t\,dt\,, 
    \nonumber\\
    & W_5(t)
    =-\frac2{105}t^3(1-t)^{11}-\frac1{20}t^6(1-t)^8
    +\frac3{70}t^8(1-t)^6-\frac1{84}t^{10}(1-t)^4\,.
\end{align}

To obtain an explicit special-function representation of the
mixed-chirality potentials, we rewrite the generating kernel
in rational form. Substituting the integral representation of $F$ [i.e., Eq.~\eqref{eq:seed}] into Eq.~\eqref{eq:potential-generator} and performing the radial integration gives
\begin{equation*}
    \frac{\Ecal}{\mathfrak a^2}
    =-\int_0^1\frac{s^2T_s^2}{B_s^2H_s}\,ds\,.
\end{equation*}
Combining this expression with the Feynman-parameter integral over $t$ yields rational double integrals. Binomial expansion followed by beta integration expresses these integrals as finite combinations of Horn-type multivariate hypergeometric functions, using techniques also developed for Feynman integrals in quantum field theory \cite{Klausen:2019hrg}. The Supplemental Information gives their explicit parameters and specifies the analytic continuation selecting the regular physical branch. The remaining theory-specific potentials are then given by
\begin{equation} \label{eq:potential-assembly}
    \Pot_{\Kinv^2}=\Pot^{\rm mix}_{4}+\Pot^{\rm ch}_{4}\,,\qquad
    \Pot_{\Pinv^2}=\Pot^{\rm mix}_{4}-\Pot^{\rm ch}_{4}\,,\qquad
    \Pot_{\Kinv\Jinv}=\Pot^{\rm mix}_{5}+\Pot^{\rm ch}_{5}\,,
\end{equation}
following the structure of Eq.~\eqref{eq:Teukolsky_sources_all} again. We observe that the single-chirality contributions to $\Pot$ follow from a finite number of derivatives of the elementary seed $\Ecal$, while mixed-chirality contributions introduce additional integrals in the Feynman parameter $t$, which combines different auxiliary values of $\mathfrak{a}$ at the same physical spin.

\subsection*{Surface gravity, horizon angular velocity, and black hole shadows}

To compute the horizon quantities from our analytic metrics, we hold the ADM mass and angular momentum fixed and normalize the stationary Killing field at infinity. The horizon is determined by requiring its generator $\xi^{\mu}=(\partial_t)^{\mu}+\omega_{+}(\partial_\phi)^{\mu}$ to be null and normal to it. The horizon angular velocity $\omega_{+}$ and surface gravity $\kappa$ then follow from
\begin{equation}
    \omega_{+}
    =-\left.\frac{g_{t\phi}}{g_{\phi\phi}}\right|_{\mathcal{H}}\,,\qquad
    \left.\nabla_\mu(\xi^{\nu}\xi_{\nu})\right|_{\mathcal{H}}
    =-2\kappa\left.\xi_{\mu}\right|_{\mathcal{H}}\,,
\end{equation}
where $\mathcal{H}$ denotes the perturbed horizon \cite{Wald:1999vt}. We retain its displacement from the Kerr horizon and the change in its generator consistently at first order. Angular independence provides an internal consistency check. For the cubic and quartic theories, we compare the resulting $\omega_{+}$ and temperature $T_{\rm H}=\kappa/(2\pi)$ with the independent thermodynamic results of~\cite{Reall:2019sah}, after converting between coupling conventions. The Supplemental Information details the calculation of the surface gravity and horizon angular velocity from our linearized metric and the analytic simplifications that reproduce the results of~\cite{Reall:2019sah}.

The shadow contours are found by backward integration of null geodesics in the reconstructed metrics. Photons are minimally coupled in the geometric-optics limit. A static observer at coordinate radius $r_{\rm obs}=50M$ defines a local screen through an orthonormal tetrad, and the figures use inclination $i=90^\circ$. Rays obey Hamilton's equations with
\begin{equation} \label{eq:null-Hamiltonian}
    H=\tfrac12g^{\mu\nu}p_\mu p_\nu=0\,.
\end{equation}
We integrate the coupled Hamilton equations directly, without assuming a separable Hamilton--Jacobi equation or a Kerr-like Carter constant. At each screen angle, capture--escape brackets are bisected in screen radius. The capture of photons is diagnosed at an inner surface near the Kerr horizon, rather than at an independently determined modified event horizon. Escaping rays return outward to the observer radius, and unresolved trajectories are excluded. 

We estimate the derivative of the shadow radius with respect to the dimensionless coupling $\lambda_I$ at $\lambda_I=0$ using symmetric positive and negative probes, while holding the background parameters and observer configuration fixed. Multiplying this derivative by the target $\lambda_I$ gives the first-order contour displacement. This isolates the first-order displacement from the finite-probe trajectories. The Kerr reference uses the same observer and screen geometry. Only the displacement is magnified for display, $\rho_{\rm shown}=\rho_{\rm Kerr}+A\,\delta\rho$, which is clearly degenerate with $\lambda_I$ because $\delta \rho \propto \lambda_I$. Details of the ray-tracing implementation using our analytic metrics are provided in the Supplemental Information.
\end{document}


\onecolumngrid
\begin{center}
{\large\bfseries Supplemental Information for}\\[4pt]
{\large\bfseries One Method to Find Them All: Analytic Black Holes beyond Einstein}\\[6pt]
\end{center}

\section{Common notation and the master potentials}

We present the potentials and source-driven metric corrections for
the six interactions
$I=(\Jinv,\Jodd,\Kinv^2,\Pinv^2,\Kinv\Pinv,\Kinv\Jinv)$,
using the conventions and fixed-charge prescription of the main
text and Methods. Their dimensionful action coefficients are
$c_I=(\alpha_{\rm e},\alpha_{\rm o},\beta_1,\beta_2,\beta_3,\gamma_1)$,
with curvature orders $n_I=(3,3,4,4,4,5)$, respectively.
For each interaction considered separately, we use
\begin{equation}\label{sm:dimensionless-normalization}
    \epsilon=\lambda_I\equiv\frac{c_I}{M^{2n_I-2}}\,,\qquad
    g_{\mu\nu}=g^{(0)}_{\mu\nu}
    +\epsilon h^{I(1)}_{\mu\nu}
    +\mathcal O(\epsilon^2)\,.
\end{equation}
All first-order fields, potentials, sources, and metric deformations
below are coefficients of the dimensionless coupling $\lambda_I$.
The displayed solutions use the prescribed fixed-charge
representative without additional homogeneous completion.

Our background notation is
\begin{align}\label{sm:notation}
    & x=\cos\theta\,,\quad w=1-x^2\,,\quad y=r-M\,,\quad
    b=\sqrt{M^2-a^2}\,,\quad \Delta=y^2-b^2\,,
    \nonumber\\
    & \Gm=r+iax\,,\quad \Gb=r-iax\,,\quad
    \Sigma=\Gm\Gb\,,\quad j=ia\sin\theta\,.
\end{align}
Tetrad components are defined by
$h_{ab}^{(1)}=e_a^\mu e_b^\nu h_{\mu\nu}^{(1)}$
in the background Kinnersley tetrad, and the angular operators are
$\mathscr L_k=\partial_\theta+k\cot\theta$.
For each interaction, the potential $\Pot$ and the three
source-driven functions $(\Ts,\Cs,\Js)$ determine the metric
perturbation. Suppressing the interaction label, the independent
components in the traceful outgoing radiation gauge are
\begin{subequations}\label{sm:metric-reconstruction}
\begin{align}
    & h_{m\bar m}^{(1)}
    =\Ts\,,\\
    & h_{\bar m\bar m}^{(1)}
    =-\frac{2\Gm}{\Gb^2}
    \left(\Gm\partial_r^2-2\partial_r\right)\Pot\,,\\
    & h_{l\bar m}^{(1)}
    =\frac{2\sqrt2\Gm}{\Gb^2\Delta}
    \left[\Sigma\mathscr L_2\partial_r
    -2r\mathscr L_2-j(\Gm-\Gb)\partial_r\right]\Pot+\Cs\,,\\
    & h_{ll}^{(1)}
    =\frac{\Sigma^2}{\Delta^2}
    \left[\Js-8\operatorname{Re}
    \left\{\frac1{\Gb^2}\left(\mathscr L_1+\frac{2j}{\Gm}\right)
    \mathscr L_2\Pot\right\}\right]\,.
\end{align}
\end{subequations}
The functions $\Ts$ and $\Js$ are real, whereas $\Cs$ and $\Pot$ can be complex. The remaining components follow from tensor symmetry and $h_{na}^{(1)}=0$, $h_{lm}^{(1)}=\bar{h}_{l\bar m}^{(1)}$, and $h_{mm}^{(1)}=\bar{h}_{\bar m\bar m}^{(1)}$. The terms containing $\Pot$ constitute $\mathscr R_{ab}[\Pot]$, while the remaining terms define the source-driven correction $\mathscr Q_{ab}[\Src]$. For all six solutions, the associated Weyl perturbations satisfy
\begin{equation}\label{sm:weyl}
    \Psi_0^{(1)}=\partial_r^4\Pot\,,\qquad
    \Psi_4^{(1)}
    =\frac{\Delta^2}{4\Gb^4}\Psi_0^{(1)}\,.
\end{equation}
We give the six potentials below, followed by their
theory-specific source-driven functions.

Using the normalized kernel $\Ecal(y,x;\mathfrak m,\mathfrak a)$ defined in the Methods section, we set $[f]_*=[f]_{\mathfrak m=M,\mathfrak a=ia}$ and $[f]_t=[f]_{\mathfrak m=M,\mathfrak a=ia(2t-1)}$. All auxiliary derivatives precede substitution, with $M,a,y$, and $x$ fixed. The master potentials $\Pot$ for all six interactions are
\begin{equation}\label{sm:potentials}
    \Pot_{\Jinv}=\Pot^{\rm ch}_3\,,\quad
    \Pot_{\Jodd}=-i\Pot^{\rm ch}_3\,,\quad
    \Pot_{\Kinv\Pinv}=-i\Pot^{\rm ch}_4\,,\quad
    \Pot_{\Kinv^2}=\Pot^{\rm mix}_4+\Pot^{\rm ch}_4\,,\quad
    \Pot_{\Pinv^2}=\Pot^{\rm mix}_4-\Pot^{\rm ch}_4\,,\quad
    \Pot_{\Kinv\Jinv}=\Pot^{\rm mix}_5+\Pot^{\rm ch}_5\,,
\end{equation}
where
\begin{subequations}
\begin{align}
    & \Pot^{\rm ch}_3
    =\frac{M^{6}w}{42}
    \left[(3+\mathfrak a\partial_{\mathfrak a})
    \partial_{\mathfrak m}^4\Ecal\right]_*\,, \label{sm:P3ch}\\
    & \Pot^{\rm ch}_4
    =-\frac{M^{9}w}{1575}
    \left[(9+2\mathfrak a\partial_{\mathfrak a})
    \partial_{\mathfrak m}^7\Ecal\right]_*\,, \label{sm:P4ch}\\
    & \Pot^{\rm ch}_5
    =\frac{M^{12}w}{655200}
    \left[(12+2\mathfrak a\partial_{\mathfrak a})
    \partial_{\mathfrak m}^{10}\Ecal\right]_*\,,\label{sm:P5ch}\\
    & \Pot^{\rm mix}_4
    =4M^{9}a^2w\int_0^1 t^3(1-t)^8
    \left[\partial_{\mathfrak m}^7
    \left(\frac{\Ecal}{\mathfrak a^2}\right)\right]_t\,dt\,,
    \label{sm:P4mix}\\
    & \Pot^{\rm mix}_5
    =M^{12}a^2w\int_0^1 W_5(t)
    \left[\partial_{\mathfrak m}^{10}
    \left(\frac{\Ecal}{\mathfrak a^2}\right)\right]_t\,dt\,,
    \label{sm:P5mix}\\
    & W_5(t)
    =-\frac2{105}t^3(1-t)^{11}-\frac1{20}t^6(1-t)^8
    +\frac3{70}t^8(1-t)^6-\frac1{84}t^{10}(1-t)^4\,.\label{sm:W5}
\end{align}    
\end{subequations}
For the mixed-chirality potentials $\Pot_4^{\rm mix}$ and
$\Pot_5^{\rm mix}$, let us define the weighted Feynman-parameter
integral
\begin{equation}\label{sm:Horn-correspondence}
    \mathcal U_{m,n}(\mathfrak m)
    =\int_0^1 t^{m-1}(1-t)^{n-1}
    \left[\frac{\Ecal}{\mathfrak a^2}\right]_{
    \mathfrak a=ia(2t-1)}dt\,,
    \qquad m,n\ge1\,.
\end{equation}
Writing
$\mathcal U_{m,n}^{[k]}
=[\partial_{\mathfrak m}^{k}\mathcal U_{m,n}]_{\mathfrak m=M}$,
the two mixed-chirality potentials are then
\begin{align}
    \Pot_4^{\rm mix}
    &=4M^9a^2w\,\mathcal U_{4,9}^{[7]}\,,
    \label{sm:Horn-P4}\\
    \Pot_5^{\rm mix}
    &=M^{12}a^2w\left[
        -\frac{2}{105}\mathcal U_{4,12}^{[10]}
        -\frac{1}{20}\mathcal U_{7,9}^{[10]}
        +\frac{3}{70}\mathcal U_{9,7}^{[10]}
        -\frac{1}{84}\mathcal U_{11,5}^{[10]}
    \right]\,.
    \label{sm:Horn-P5}
\end{align}

A solution expressed in quadrature is not as useful as one where the solution to the quadrature is represented in closed form. Let us then now provide an equivalent Horn-type hypergeometric
representation of $\mathcal U_{m,n}$ \cite{Klausen:2019hrg}. After substituting $\mathfrak a=ia(2t-1)$ in the generating
kernel $\mathcal{E}$, we write its polynomials as functions of the two
independent integration parameters $s,t\in[0,1]$:
\begin{align} \label{sm:kernel-polynomials}
    & H(s,t)=B(s,t)[y+b+s\,d(t)]+ia(2t-1)sT(s,t)(x-1)\,,\qquad
    B(s,t)=\mathfrak m+b+ia(2t-1)(2s-1)\,, \nonumber\\
    & T(s,t)=2b+s\,d(t)\,,\qquad
    d(t)=\mathfrak m+ia(2t-1)-b\,.
\end{align}
The auxiliary parameter $\mathfrak m$ remains independent. The rational kernel representation becomes
\begin{equation}\label{sm:Horn-rational-kernel}
    \left[\frac{\Ecal}{\mathfrak a^2}\right]
    _{\mathfrak a=ia(2t-1)}
    =-\int_0^1\frac{s^2T(s,t)^2}{B(s,t)^2H(s,t)}\,ds\,.
\end{equation}
We normalize the denominator of the polynomials by their constant terms in both integration parameters:
\begin{equation} \label{sm:Horn-normalization}
    B_{00}\equiv B(0,0)=\mathfrak m+b+ia\,,\qquad
    H_{00}\equiv H(0,0)=(y+b)B_{00}\,.
\end{equation}
The ten dimensionless arguments $\boldsymbol z=(z_1,\ldots,z_{10})$ and the numerator coefficients $d_{[u,v]}$ are defined by
\begin{align} \label{sm:Horn-arguments}
    & \frac{B(s,t)}{B_{00}}
    =1-\sum_{h=1}^{3}z_hs^{p_h}t^{q_h}\,,\qquad
    \frac{H(s,t)}{H_{00}}
    =1-\sum_{h=4}^{10}z_hs^{p_h}t^{q_h}\,,\qquad
    s^2T(s,t)^2
    =\sum_{u=2}^{4}\sum_{v=0}^{u-2}d_{[u,v]}s^ut^v\,,\nonumber\\
    & (p_h,q_h)
    =\big((1,0),(0,1),(1,1);\,
    (0,1),(1,0),(1,1),(1,2),(2,0),(2,1),(2,2)\big)\,.
\end{align}
Explicitly,
\begin{align} \label{sm:Horn-argument-values}
    & z_1=z_2=z_4=\frac{2ia}{B_{00}}\,,\quad
    z_3=-\frac{4ia}{B_{00}}\,,\quad
    z_5=\frac{b^2-\mathfrak m^2-a^2
               +2ia[y+b(1+x)]}{H_{00}}\,,\quad
    z_6=\frac{4a^2-4ia[y+b(1+x)]}{H_{00}}\,,\nonumber\\
    & z_7=-\frac{4a^2}{H_{00}}\,,\quad
    z_8=\frac{ia(1+x)(\mathfrak m-b-ia)}{H_{00}}\,,\quad
    z_9=-\frac{2ia(1+x)(\mathfrak m-b-2ia)}{H_{00}}\,,\quad
    z_{10}=\frac{4a^2(1+x)}{H_{00}}\,.
\end{align}
The six potentially nonzero numerator coefficients are
\begin{align} \label{sm:Horn-numerator-coefficients}
    & d_{[2,0]}=4b^2\,,\quad
    d_{[3,0]}=4b(\mathfrak m-b-ia)\,,\quad
    d_{[3,1]}=8iab\,,\nonumber\\
    & d_{[4,0]}=(\mathfrak m-b-ia)^2\,,\quad
    d_{[4,1]}=4ia(\mathfrak m-b-ia)\,,\quad
    d_{[4,2]}=-4a^2\,,
\end{align}
while other coefficients $d_{[u,v]}$ vanish, so $u\in\{2,3,4\}$ and $v\in\{0,1,2\}$.

Introduce the summation multi-index
$\boldsymbol k=(k_1,\ldots,k_{10})\in\mathbb N_0^{10}$,
where each $k_h$ is an independent nonnegative integer,
and define
\begin{equation}\label{sm:Horn-index-sums}
    N_B=\sum_{h=1}^{3}k_h\,,\qquad
    N_H=\sum_{h=4}^{10}k_h\,,\qquad
    N_s=\sum_{h=1}^{10}p_hk_h\,,\qquad
    N_t=\sum_{h=1}^{10}q_hk_h\,.
\end{equation}
We define the required Horn-type hypergeometric
function \cite{Klausen:2019hrg} by
\begin{equation}\label{sm:Horn-series}
    \Phi_{\rho,\sigma;\alpha,\beta,\delta}(\boldsymbol z)
    =\sum_{\boldsymbol k\in\mathbb N_0^{10}}
    \frac{(\rho)_{N_B}(\sigma)_{N_H}
    (\alpha)_{N_s}(\beta)_{N_t}}
    {(\alpha+1)_{N_s}(\beta+\delta)_{N_t}}
    \prod_{h=1}^{10}\frac{z_h^{k_h}}{k_h!}\,,
\end{equation}
where $(c)_N$ denotes the rising factorial. The weighted
integral in Eq.~\eqref{sm:Horn-correspondence} is then
\begin{equation}\label{sm:Horn-master}
    \mathcal U_{m,n}(\mathfrak m)
    =-\frac{1}{B_{00}^{\,2}H_{00}}
    \sum_{u=2}^{4}\sum_{v=0}^{u-2}
    \frac{d_{[u,v]}\,\mathrm B(m+v,n)}{u+1}
    \Phi_{2,1;u+1,m+v,n}(\boldsymbol z)\,,
\end{equation}
where $\mathrm B(\cdot)$ is the Euler beta function. Substituting
this expression into Eqs.~\eqref{sm:Horn-P4}
and~\eqref{sm:Horn-P5} gives the hypergeometric
representations of both mixed-chirality potentials.

Equation~\eqref{sm:Horn-master} is a finite combination of six
hypergeometric functions. Their defining series generally do not terminate, but the indices $\boldsymbol k$ label powers of the special-function arguments, not angular multipoles; no angular truncation is involved. For the parameters used here, a sufficient region of absolute convergence is $\sum_{h=1}^{3}|z_h|<1$ and $\sum_{h=4}^{10}|z_h|<1$. Outside the series' convergence domain, $\Phi$ is defined by analytic continuation selected by the rational double-integral representation in Eqs.~\eqref{sm:Horn-correspondence} and~\eqref{sm:Horn-rational-kernel}. This selects the regular physical branch, since at $\mathfrak m=M$ both $\operatorname{Re}B(s,t)$ and $\operatorname{Re}H(s,t)$ remain positive throughout the subextremal exterior for $s,t\in[0,1]$. All auxiliary derivatives act on the continued expression, including $B_{00}$, $H_{00}$, $d_{[u,v]}$, and $\boldsymbol z$, at fixed $M,a,y,x$, before setting $\mathfrak m=M$.

\section{Source-driven metric corrections}

Define the symmetric and antisymmetric Laurent combinations
\begin{equation}\label{sm:laurent}
    S_{[m,n]}=\Gm^{-m}\Gb^{-n}+\Gm^{-n}\Gb^{-m}\,,\qquad
    D_{[m,n]}=\Gm^{-m}\Gb^{-n}-\Gm^{-n}\Gb^{-m}\,,\qquad 
    m,n\in\mathbb Z\,.
\end{equation}
For the polynomial expressions below, we use
\begin{equation}\label{sm:ratios}
    z=\frac{\Gm}{\Gb}\,,\qquad 
    \widehat M=\frac{M}{\Gb}\,,\qquad
    \widehat a=\frac{a}{\Gb}\,,\qquad
    \Lambda_R=\log\frac{\Gm}{\Gb}=2i\arctan\frac{ax}{r}\,.
\end{equation}
Apparent singularities at $z=1$ are removable: the rational and logarithmic terms must be combined before taking the limit $ax\to0$.

\begin{table}[!t]
    \centering
    \footnotesize
    \setlength{\tabcolsep}{2pt}
    \renewcommand{\arraystretch}{1.2}
    \begin{tabular}{@{}l*{14}{r}@{}}
    \toprule
    Index $j$
    & $0$ & $1$ & $2$ & $3$ & $4$ & $5$ & $6$
    & $7$ & $8$ & $9$ & $10$ & $11$ & $12$ & $13$\\
    \midrule
    $c_j^{T,4}$
    & $24$ & $-245$ & $1200$ & $-3920$ & $11000$ & $-7718$ & ---
    & --- & --- & --- & --- & --- & --- & ---\\
    $c_j^{M,4}$
    & $-176$ & $1732$ & $-7632$ & $19912$ & $-37216$ & $-35908$ & $-38944$
    & $196464$ & --- & --- & --- & --- & --- & ---\\
    $c_j^{A,4}$
    & $416$ & $-5476$ & $36884$ & $-187804$ & $1097100$ & $7704088$ & $11010792$
    & --- & --- & --- & --- & --- & --- & ---\\
    $c_j^{0,4}$
    & $104$ & $-1297$ & $8623$ & $-45162$ & $265398$ & $1132521$ & $-653351$
    & $-706836$ & --- & --- & --- & --- & --- & ---\\
    $g_j^{(4)}$
    & $16$ & $-155$ & $643$ & $-1309$ & $-595$ & $-7402$ & $25190$
    & $-24986$ & $340578$ & $-21455$ & $-69889$ & $13215$ & $-2007$ & $156$\\
    \bottomrule
    \end{tabular}
    \caption{Quartic polynomial coefficients. Dashes indicate
    entries not required by the polynomial definitions,
    rather than zero coefficients.}
    \label{tab:quartic-coefficients}
\end{table}

\subsection{Cubic interactions}

For the parity-even cubic interaction $\Jinv$,
\begin{subequations}
\begin{align}
    \Ts_{\Jinv}
    &=-\frac{108M^{6}}{35}\left(6S_{[6,0]}+S_{[5,1]}\right)\,,\label{sm:Te}\\
    \Cs_{\Jinv}
    &=\frac{\sqrt2j\Gm}{\Delta}\frac{M^{6}}{35}
    \left(\frac{368}{\Gm^6}+\frac{348}{\Gm^5\Gb}
    +\frac{40}{\Gm^4\Gb^2}+\frac{108}{\Gm\Gb^5}
    +\frac{648}{\Gb^6}\right)\,,\label{sm:Ce}\\
    \Js_{\Jinv}
    &=\frac{16M^{7}}{35}
    \left(-8S_{[7,0]}-5S_{[6,1]}+27S_{[5,2]}\right)
    -\frac{16M^{6}a^2}{35}
    \left(92wS_{[7,1]}+(87-67x^2)S_{[6,2]}\right)\,.\label{sm:Je}
\end{align}    
\end{subequations}
For the parity-odd cubic interaction $\Jodd$,
\begin{subequations}
\begin{align}
    \Ts_{\Jodd}
    &=\frac{108iM^{6}}{35}
    \left(6D_{[6,0]}+D_{[5,1]}\right)\,,\label{sm:To}\\
    \Cs_{\Jodd}
    &=-\frac{\sqrt2j\Gm}{\Delta}\frac{iM^{6}}{35}
    \left(\frac{368}{\Gm^6}+\frac{348}{\Gm^5\Gb}
    +\frac{40}{\Gm^4\Gb^2}-\frac{108}{\Gm\Gb^5}
    -\frac{648}{\Gb^6}\right)\,,\label{sm:Co}\\
    \Js_{\Jodd}
    &=-\frac{16iM^{7}}{35}
    \left(-8D_{[7,0]}-5D_{[6,1]}+27D_{[5,2]}\right)
    +\frac{16iM^{6}a^2}{35}
    \left(92wD_{[7,1]}+(87-67x^2)D_{[6,2]}\right)\,.
    \label{sm:Jo}
\end{align}
\end{subequations}

\subsection{Parity-odd quartic interaction}

For the parity-odd quartic interaction $\Kinv\Pinv$,
\begin{subequations}
\begin{align}
    \Ts_{\Kinv\Pinv}
    =& \;\frac{56iM^{9}}{5}
    \left(9D_{[9,0]}+D_{[8,1]}\right)\,,\label{sm:TKQ}\\
    \Cs_{\Kinv\Pinv}
    =& \;\frac{\sqrt2j\Gm}{\Delta}\frac{8iM^{9}}{5}
    \left(\frac{63}{\Gb^9}+\frac7{\Gm\Gb^8}-\frac8{\Gm^7\Gb^2}
    -\frac{79}{\Gm^8\Gb}-\frac{303}{\Gm^9}\right)\,,
    \label{sm:CKQ}\\
    \Js_{\Kinv\Pinv}
    =& \;\frac{8iM^{9}}{5}
    \left[-M(508D_{[10,0]}+624D_{[9,1]}+28D_{[8,2]})
    +a^2(1212D_{[10,1]}+316D_{[9,2]})\right. \nonumber\\
    & \;\left.+303D_{[10,-1]}+97D_{[9,0]}+177D_{[8,1]}+63D_{[7,2]}\right]\,.
    \label{sm:JKQ}
\end{align}    
\end{subequations}

\subsection{Parity-even quartic interactions}

For the parity-even quartic interactions, $I=\Kinv^2$ or $\Pinv^2$, with $\sigma_{\Kinv^2}=1$ and $\sigma_{\Pinv^2}=-1$,
\begin{equation} \label{sm:quartic-split}
    \Ts_I
    =\frac{M^{9}}{\Gb^9}c_{\Ts}+\sigma_I d_{\Ts}\,,\qquad
    \Cs_I
    =\frac{\sqrt2j\Gm}{\Delta}
    \left(\frac{M^{9}}{\Gb^9}c_{\Cs}+\sigma_I d_{\Cs}\right)\,,\qquad
    \Js_I
    =\frac{M^{9}}{\Gb^9}c_{\Js}+\sigma_I d_{\Js}\,,
\end{equation}
where
\begin{align}
    c_{\Ts}
    =& \;-\frac{48(z+1)\Pi_T^{(4)}(z)}{5z^6(z-1)^8}
    +\frac{80640(z^2+1)\Lambda_R}{z(z-1)^9}\,, 
    \nonumber \\
    c_{\Cs}
    =& \;-\frac{48\Pi_C^{(4)}(z)}{5z^6(z-1)^{10}}
    -\frac{80640L_C^{(4)}(z)\Lambda_R}{z(z-1)^{11}}\,, 
    \nonumber \\
    c_{\Js}
    =& \;-\frac{48[\widehat M\Pi_M^{(4)}(z)
    +\widehat a^2\Pi_A^{(4)}(z)+\Pi_0^{(4)}(z)]}{5z^7(z-1)^{10}}
    +\frac{80640[\widehat M L_M^{(4)}(z)
    +\widehat a^2L_A^{(4)}(z)+L_0^{(4)}(z)]\Lambda_R}
    {z^2(z-1)^{11}}\,, \nonumber\\
    d_{\Ts}
    =& \;-\frac{56M^{9}}{5}\left(9S_{[9,0]}+S_{[8,1]}\right)\,,
    \nonumber\\
    d_{\Cs}
    =& \;\frac{8M^{9}}{5}\left(\frac{63}{\Gb^9}+\frac7{\Gm\Gb^8}
    +\frac8{\Gm^7\Gb^2}+\frac{79}{\Gm^8\Gb}
    +\frac{303}{\Gm^9}\right)\,, \nonumber\\
    d_{\Js}
    =& \;\frac{8M^{9}}{5}\left[M\left(508S_{[10,0]}+624S_{[9,1]}+28S_{[8,2]}\right)
    -a^2\left(1212S_{[10,1]}+316S_{[9,2]}\right)\right. \nonumber\\
    & \;\left.-303S_{[10,-1]}-97S_{[9,0]}-177S_{[8,1]}-63S_{[7,2]}\right]\,.
\end{align}
\begin{table}[!t]
    \centering
    \footnotesize
    \setlength{\tabcolsep}{3pt}
    \renewcommand{\arraystretch}{1.2}
    \begin{tabular}{@{}l*{7}{r}@{}}
    \toprule
    Index $j$
    & $0$ & $1$ & $2$ & $3$ & $4$ & $5$ & $6$\\
    \midrule
    $c_j^{T,5}$
    & $128520$ & $-1274490$ & $5676300$ & $-14895405$
    & $25222090$ & $-27768430$ & $17392818$\\
    $c_j^{M,5}$
    & $-283080$ & $3045840$ & $-14512680$ & $39277420$
    & $-61796020$ & $39585260$ & $56598828$\\
    $c_j^{A,5}$
    & $994560$ & $-11670120$ & $62465760$ & $-200518780$
    & $422746720$ & $-581699580$ & $390357792$\\
    $c_j^{0,5}$
    & $248640$ & $-3063690$ & $17505180$ & $-61337185$
    & $146104270$ & $-243793550$ & $264261998$\\
    \midrule
    Index $j$
    & $7$ & $8$ & $9$ & $10$ & $11$ & $12$ & $13$\\
    \midrule
    $c_j^{T,5}$
    & $-719144$ & $-7467754$ & $-4306464$ & $46422346$
    & $116332186$ & --- & ---\\
    $c_j^{M,5}$
    & $-181652328$ & $216504288$ & $-45813768$ & $-434113680$
    & $530122120$ & $-146962200$ & ---\\
    $c_j^{A,5}$
    & $457415640$ & $-2282936832$ & $6057804720$ & $-18806177280$
    & $-53712842920$ & $-46251156160$ & ---\\
    $c_j^{0,5}$
    & $-65735241$ & $-547515540$ & $1996940946$ & $-6312954648$
    & $-2888657750$ & $9654755220$ & $-3913517300$\\
    \bottomrule
    \end{tabular}
    \caption{Quintic polynomial coefficients. Dashes indicate
    entries not required by the polynomial definitions,
    rather than zero coefficients.}
    \label{tab:quintic-coefficients}
\end{table}
Here
\begin{align} \label{sm:quartic-logs}
    L_C^{(4)}&=14z^5-27z^4-18z^3+2z^2-2z+1\,, \nonumber\\
    L_M^{(4)}&=-4z^5-8z^4+12z^3+12z^2-8z-4\,, \nonumber\\
    L_A^{(4)}&=508(z^4+1)+1168(z^3+z)+1328z^2\,, \nonumber\\
    L_0^{(4)}&=99(z^6+1)+42(z^5+z)-109(z^4+z^2)-64z^3\,.
\end{align}
Define the palindromic-polynomial notation
\begin{equation}\label{sm:palindrome}
    \Pol_N[c_0,\ldots,c_{\lfloor N/2\rfloor}](z)
    =\sum_{j=0}^{\lfloor N/2\rfloor}
    c_j\frac{z^j+z^{N-j}}{1+\delta_{2j,N}}\,.
\end{equation}
Then
\begin{equation} \label{sm:quartic-polys}
    \Pi_T^{(4)}=\Pol_{10}[c^{T,4}]\,,\qquad
    \Pi_M^{(4)}=\Pol_{14}[c^{M,4}]\,,\qquad
    \Pi_A^{(4)}=\Pol_{13}[c^{A,4}]\,,\qquad
    \Pi_0^{(4)}=\Pol_{15}[c^{0,4}]\,,\qquad
    \Pi_C^{(4)}=\sum_{j=0}^{13}g_j^{(4)}z^j\,,
\end{equation}
with coefficients in Table~\ref{tab:quartic-coefficients}.

\subsection{Quintic interaction}

For the quintic interaction $\Kinv\Jinv$,
\begin{subequations}
    \begin{align}
    \Ts_{\Kinv\Jinv}
    &=\frac{M^{12}}{\Gb^{12}}
    \left[\frac{463104(z+1)\Lambda_R}{z(z-1)^{11}}
    -\frac{24\Pi_T^{(5)}(z)}{5005z^{12}(z-1)^{10}}\right]\,,\label{sm:T5}\\
    \Cs_{\Kinv\Jinv}
    &=\frac{\sqrt2j\Gm}{\Delta}
    \frac{M^{12}}{\Gb^{12}}
    \left[\frac{6912(605z^3-67z^2+869z+67)\Lambda_R}{z(z-1)^{12}}
    +\frac{24\Pi_C^{(5)}(z)}{5005z^{12}(z-1)^{11}}\right]\,,
    \label{sm:C5} \\
    \Js_{\Kinv\Jinv}
    &=-\frac{M^{12}}{\Gb^{12}}
    \left[\frac{6912[\widehat M L_M^{(5)}(z)
    +\widehat a^2L_A^{(5)}(z)+L_0^{(5)}(z)]\Lambda_R}{z^2(z-1)^{13}}
    +\frac{24[\widehat M\Pi_M^{(5)}(z)
    +\widehat a^2\Pi_A^{(5)}(z)+\Pi_0^{(5)}(z)]}
    {5005z^{13}(z-1)^{12}}\right]\,,\label{sm:J5}
\end{align}   
\end{subequations}
where
\begin{align} \label{sm:quintic-logs}
    L_M^{(5)}&=268(z-1)^2(z^2+1)\,,\nonumber\\
    L_A^{(5)}&=28772(z^3+1)+34588(z^2+z)\,,\nonumber\\
    L_0^{(5)}&=5983(z^5+1)-4749(z^4+z)-1234(z^3+z^2)\,,
\end{align}
and
\begin{equation} \label{sm:quintic-polys}
    \Pi_T^{(5)}=\Pol_{22}[c^{T,5}]\,,\qquad
    \Pi_M^{(5)}=\Pol_{25}[c^{M,5}]\,,\qquad
    \Pi_A^{(5)}=\Pol_{24}[c^{A,5}]\,,\qquad
    \Pi_0^{(5)}=\Pol_{26}[c^{0,5}]\,,\qquad
    \Pi_C^{(5)}=\sum_{j=0}^{23}g_j^{(5)}z^j\,,
\end{equation}
with coefficients in Tables~\ref{tab:quintic-coefficients} and \ref{tab:quintic-g-coefficients}.

\begin{table}[!t]
    \centering
    \footnotesize
    \setlength{\tabcolsep}{3pt}
    \renewcommand{\arraystretch}{1.2}
    \begin{tabular}{@{}l*{8}{r}@{}}
    \toprule
    Index $j$
    & $0$ & $1$ & $2$ & $3$ & $4$ & $5$ & $6$ & $7$\\
    \midrule
    $g_j^{(5)}$
    & $-248640$ & $2668890$ & $-12952170$ & $37473205$
    & $-71774505$ & $95704950$ & $-91355638$ & $63421556$\\
    \midrule
    Index $j$
    & $8$ & $9$ & $10$ & $11$ & $12$ & $13$ & $14$ & $15$\\
    \midrule
    $g_j^{(5)}$
    & $-28427916$ & $-17105176$ & $117738104$ & $-773708606$
    & $-841913886$ & $90255484$ & $-1005140576$ & $468477284$\\
    \midrule
    Index $j$
    & $16$ & $17$ & $18$ & $19$ & $20$ & $21$ & $22$ & $23$\\
    \midrule
    $g_j^{(5)}$
    & $-239032948$ & $127850998$ & $-75351430$ & $43942745$
    & $-20880585$ & $6950790$ & $-1403010$ & $128520$\\
    \bottomrule
    \end{tabular}
    \caption{Complete coefficient sequence $g_j^{(5)}$
    for the quintic polynomial.}
    \label{tab:quintic-g-coefficients}
\end{table}

\section{Shift of the extremality bound}
\label{sm:extremality}

Let us distinguish Kerr extremality, $\chi=a/M=1$, from the corrected zero-temperature relation $\chi_{\rm ext}=J_{\rm ext}(M)/M^2$. Here, $M$ and $J=aM$ are the ADM mass and angular momentum, $G=1=c$, and $J>0$; the mass formula below extends to either spin orientation by replacing $J$ with $|J|$. We retain only terms linear in the physical couplings. The deformed metric functions above use $\epsilon=\lambda_I$, while the thermodynamic action is written in terms of the dimensionful $c_I$. Thermodynamic derivatives hold $c_I$ fixed, not $\lambda_I=c_I/M^{2n_I-2}$.

Following Ref.~\cite{Reall:2019sah}, we expand the on-shell Euclidean action $I_E$ about the zero-coupling Kerr solution $g^{\rm K}(T_{\rm H},\omega_+)$, holding the Hawking temperature $T_{\rm H}$ and horizon angular velocity $\omega_+$ fixed:
\begin{equation}\label{sm:action-expansion}
    I_E(T_{\rm H},\omega_+;\{c_I\})
    =I_E^{\rm K}(T_{\rm H},\omega_+)
    +\delta I_E+\mathcal O(c^2)\,,\quad
    \delta I_E
    =\sum_I c_I\left.\frac{\partial I_E}{\partial c_I}
    \right|_{\{c_J\}=0}\,.
\end{equation}
The derivatives are taken along the family of black-hole solutions at fixed $T_{\rm H}$ and $\omega_+$, so they include both the explicit coupling dependence of the
action and the induced change in the metric. The reference Kerr solution shares these thermodynamic variables with the corrected black hole, but not necessarily its mass and angular momentum.

The metric-induced variation of the Einstein action vanishes under these fixed-$T_{\rm H},\omega_+$ conditions when the Einstein boundary term and subtraction at infinity are included \cite{Reall:2019sah}. The first-order correction therefore comes entirely from the explicit higher-curvature terms evaluated on Kerr; their boundary contributions vanish at infinity for the interactions considered here. Using $I_E=\mathscr G/T_{\rm H}$, with $\mathscr G=M-T_{\rm H}S_{\rm W}-\omega_+J$ and $S_{\rm W}$ the Wald entropy, we obtain
\begin{equation}\label{sm:ext-gibbs}
    \delta\mathscr G(T_{\rm H},\omega_+)
    =T_{\rm H}\,\delta I_E
    =-\frac18\sum_I c_I
    \int_{r_+^{\rm K}}^\infty dr\int_{-1}^{1}dx\,
    \left.\Gm\Gb\,\mathcal O_I\right|_{
    g^{\rm K}(T_{\rm H},\omega_+)}\,,
\end{equation}
where $\mathcal O_I$ is the curvature interaction multiplying $c_I$ in the action of Eq.~(6) in the main text, and $r_+^{\rm K}$ is the horizon radius of the reference Kerr solution. In units $G=1$, the factor $1/8$ follows from the action normalization $1/(16\pi)$ and the azimuthal integral $2\pi$; the Euclidean-time period $1/T_{\rm H}$ cancels against the prefactor $T_{\rm H}$.

To determine the extremality bound, we next take
$T_{\rm H}\to0$ and compare extremal black holes at
the same (positive) angular momentum $J$.
The reference extremal Kerr black hole has
\begin{equation}\label{sm:ext-Kerr-reference}
    M_0\equiv M_{\rm ext}^{\rm Kerr}(J)=\sqrt J\,,
    \qquad
    \omega_0\equiv\omega_+^{\rm Kerr,ext}(J)
    =\frac{1}{2\sqrt J}\,.
\end{equation}
The subscript $0$ denotes the zero-coupling Kerr values
at this fixed $J$. These quantities specify the background
on which the first-order action correction is evaluated.

\begin{figure}[t]
    \centering
    \includegraphics[width=0.6\linewidth]{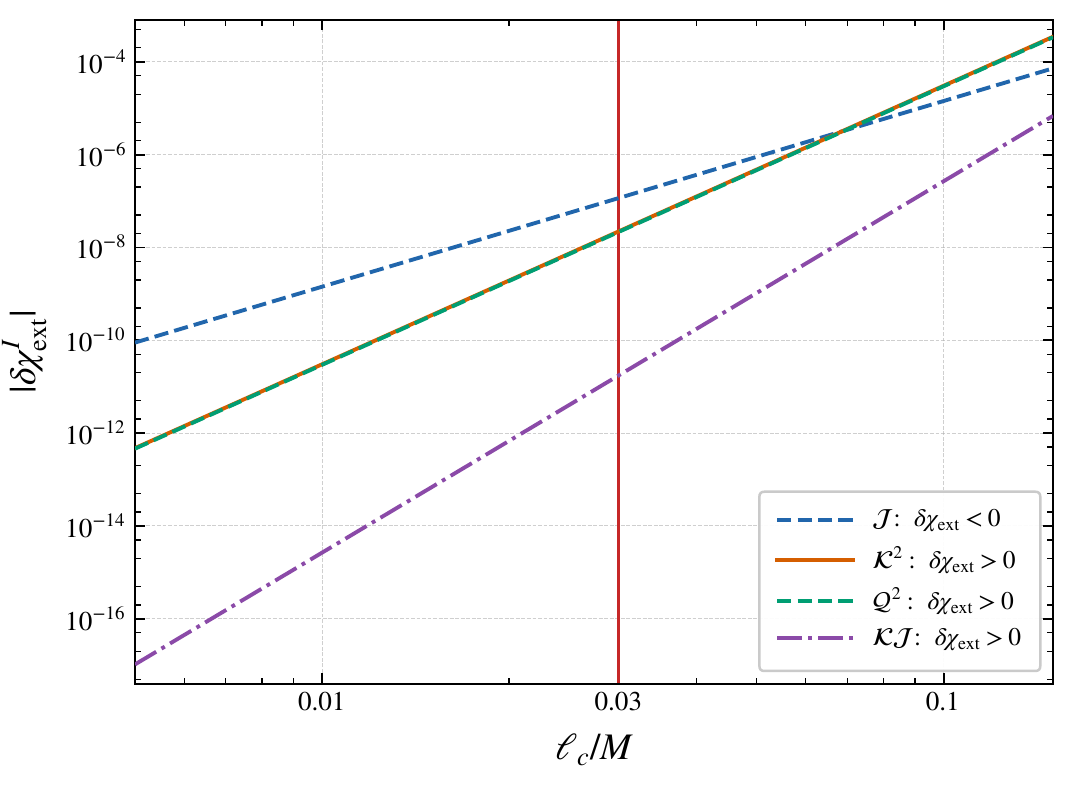}
    \caption{First-order shifts of the extremal-spin bound $\delta\chi_{\rm ext}^{I}$ at a common higher-curvature cutoff length scale $\ell_c$. Curves show $|\delta\chi_{\rm ext}^{I}|$, where the signs of the shifts are indicated in the legend. The red vertical line marks $\ell_c/M=0.03$, the common cutoff used in the main text. The parity-odd interactions $\Jodd$ and $\Kinv\Pinv$ give vanishing first-order extremality shifts and are omitted from the logarithmic plot.}
    \label{fig:extremality}
\end{figure}

At zero temperature, $M_{\rm ext}=\mathscr G+\omega_+J$, so let us write $\mathscr G=\mathscr G_{\rm K}+\delta\mathscr G$ and $\omega_+=\omega_0+\delta\omega_+$, where $\mathscr G_{\rm K}$ is the Kerr Gibbs free energy. Expanding at fixed $J$ gives
\begin{equation} \label{sm:ext-Legendre}
    \delta M_{\rm ext}(J)
    =\delta\mathscr G(0,\omega_0)
    +\left(\left.\frac{\partial\mathscr G_{\rm K}}
    {\partial\omega_+}\right|_
    {T_{\rm H}=0,\,\omega_+=\omega_0}
    +J\right)\delta\omega_+
    =\delta\mathscr G(0,\omega_0)\,,
\end{equation}
because the Kerr thermodynamic identity $\partial_{\omega_+}\mathscr G_{\rm K}=-J$ makes the parenthesis vanish. Thus, the angular-velocity correction need not be calculated to obtain the first-order extremal mass shift. Since the reference extremal Kerr solution has $a=r_+^{\rm K}=M_0$, Eq.~\eqref{sm:ext-gibbs} yields
\begin{equation}\label{sm:ext-mass-integral}
    \delta M_{\rm ext}(J)
    =-\frac18\sum_I c_I
    \int_{M_0}^\infty dr\int_{-1}^{1}dx\,
    \left.\Gm\Gb\,\mathcal O_I\right|_{M=a=M_0}\,.
\end{equation}
Here $\delta M_{\rm ext}=M_{\rm ext}-\sqrt J$ denotes the first-order change at fixed angular momentum. The calculation takes the zero-temperature limit of the thermodynamic correction, and it does not set $a=M$ in the fixed-subextremal-spin metric perturbation. The parity-odd integrands are odd under $x\mapsto-x$, so $\Jodd$ and $\Kinv\Pinv$ give no linear extremal-mass correction.

Let us now evaluate the first-order change to the extremal mass for the different higher-curvature interactions. For the quintic term, the Kerr identities
\begin{equation} \label{sm:ext-invariants}
    \Kinv=24M_0^2S_{[6,0]}\,,\qquad
    \Jinv=48M_0^3S_{[9,0]}\,,\qquad
    \Kinv\Jinv=1152M_0^5\left(S_{[15,0]}+S_{[9,6]}\right)
\end{equation}
reduce Eq.~\eqref{sm:ext-mass-integral} to an elementary rational integral. With $R=r/M_0$ and $u=x/R$, reversing the integration order gives
\begin{align} \label{sm:ext-KJ-integral}
    \int_{M_0}^\infty\!\dd r\int_{-1}^{1}\,dx\,\Sigma\Kinv\Jinv
    =\frac1{M_0^7}\left(\frac{696}{13}+\frac{135\pi}{8}\right)\,.
\end{align}
Combining all six interactions yields
\begin{align} \label{sm:ext-mass}
    M_{\rm ext}(J)
    =\sqrt{J}+\frac{\alpha_{\rm e}}{14J^{3/2}}
    -\frac{76\beta_1+74\beta_2}{5J^{5/2}}
    -\left(\frac{87}{13}+\frac{135\pi}{64}\right)
    \frac{\gamma_1}{J^{7/2}}+\mathcal O(c^2)\,, 
\end{align}
which gives the extremal mass at fixed angular momentum $J$. Inverting this relation gives the extremal angular momentum $J_{\rm ext}(M)$ at fixed ADM mass $M$, and hence
\begin{align} \label{sm:ext-spin}
    \chi_{\rm ext}(M)
    \equiv\frac{J_{\rm ext}(M)}{M^2}
    =1-\frac{\lambda_{\Jinv}}{7}
    +\frac{152\lambda_{\Kinv^2}+148\lambda_{\Pinv^2}}{5}
    +\left(\frac{174}{13}+\frac{135\pi}{32}\right)\lambda_{\Kinv\Jinv}
    +\mathcal O(\lambda^2)\,. 
\end{align}
Here $\lambda_I=c_I/M^{2n_I-2}$ is evaluated at the fixed ADM mass in Eq.~\eqref{sm:ext-spin}, and the $\mathcal O(c^2)$ and $\mathcal O(\lambda^2)$ symbols denote terms at least quadratic in the dimensionful and dimensionless couplings, respectively. The cubic and quartic terms reproduce~\cite{Reall:2019sah} after matching the dual convention: its dual Riemann tensor omits our factor $1/2$, so its squared parity-odd invariant is $4\Pinv^2$ and its coefficient $L^6\widetilde\lambda_{\rm e}$ equals $\beta_2/4$. The quintic coefficient follows from Eq.~\eqref{sm:ext-KJ-integral} and is an extension of the thermodynamic calculation in \cite{Reall:2019sah}.

The first-order shift $\delta\chi_{\rm ext}^{I}$ measures the departure of the extremal spin from the Kerr bound at fixed ADM mass: a positive shift raises the bound, whereas a negative shift lowers it. For the figures presented in the main text, each interaction is considered separately, with its dimensionless operator coefficient set to unity, so that $\lambda_I=c_I/M^{2n_I-2}=(\ell_c/M)^{2n_I-2}$.  Figure~\ref{fig:extremality} shows $|\delta\chi_{\rm ext}^{I}|$ as a function of the common cutoff $\ell_c/M$, with the red, vertical line marking the $\ell_c/M=0.03$ value used in the main text. The parity-odd interactions $\Jodd$ and $\Kinv\Pinv$ have vanishing first-order shifts and are therefore omitted from the logarithmic plot. At  $\ell_c/M=0.03$, the largest first-order shift among the six interactions occurs for parity-even cubic gravity, with $\delta\chi_{\rm ext}\simeq-1.16\times10^{-7}$. This is only about $0.12\%$ of the Kerr extremality gap $1-\chi=10^{-4}$ for the main-text examples with $a/M=0.9999$. These examples therefore remain near-extremal and below the corrected extremal bound in all six theories, at the order considered.

\bibliographystyle{apsrev4-2}
\bibliography{references}